\documentclass[apj]{emulateapj}
\usepackage{amssymb}
\usepackage{amsmath}
\usepackage{cancel}
\usepackage{bbold}
\usepackage{graphicx}
\usepackage{xspace}
\usepackage{units}
\usepackage[breaklinks,colorlinks,urlcolor=blue,citecolor=blue,linkcolor=blue]{hyperref}

\let\oldsqrt\sqrt
\def\sqrt{\mathpalette\DHLhksqrt}
\def\DHLhksqrt#1#2{
	\setbox0=\hbox{$#1\oldsqrt{#2\,}$}
	\dimen0=\ht0
	\advance\dimen0-0.2\ht0
	\setbox2=\hbox{\vrule height\ht0 depth -\dimen0}
	{\box0\lower0.4pt\box2}
}

\newcommand{\bo}{\raise-0.4pt\hbox{\large$\Box$}}			
\renewcommand{\d}{\ensuremath{\mathrm{d}}}					
\renewcommand{\arg}[1]{\ensuremath{\! \left( #1 \right)}}	

\newcommand{\pc}{\ensuremath{\, \mathrm{pc}}\xspace}
\newcommand{\kpc}{\ensuremath{\, \mathrm{kpc}}\xspace}

\newcommand{\FeH}{\ensuremath{\left[ \nicefrac{\mathrm{Fe}}{\mathrm{H}} \right]}}
\newcommand{\EBV}{\ensuremath{\mathrm{E} \! \left( B \! - \! V \right)}}
\newcommand{\mobs}{\ensuremath{\vec{m}_{\mathrm{obs}}}}

\begin{document}

\title{Measuring Distances and Reddenings for a Billion Stars: Towards A 3D Dust Map from Pan-STARRS 1}

\author{Gregory~Maurice~Green\altaffilmark{1},
Edward~F.~Schlafly\altaffilmark{2},
Douglas~P.~Finkbeiner\altaffilmark{1},
Mario~Juri\'{c}\altaffilmark{3},
Hans-Walter~Rix\altaffilmark{2},
Will~Burgett\altaffilmark{5},
Kenneth~C.~Chambers\altaffilmark{5},
Peter~W.~Draper\altaffilmark{6},
Heather~Flewelling\altaffilmark{5},
Rolf~Peter~Kudritzki\altaffilmark{5},
Eugene~Magnier\altaffilmark{5},
Nicolas~Martin\altaffilmark{4},
Nigel~Metcalfe\altaffilmark{6},
John~Tonry\altaffilmark{5},
Richard~Wainscoat\altaffilmark{5},
Christopher~Waters\altaffilmark{5}
}

\email{ggreen@cfa.harvard.edu}

\altaffiltext{1}{Harvard-Smithsonian Center for Astrophysics, 60 Garden St., Cambridge, MA 02138}
\altaffiltext{2}{Max-Planck-Institut f\"{u}r Astronomie, K\"{o}nigstuhl 17, D-69117 Heidelberg}
\altaffiltext{3}{LSST Corporation, 933 N. Cherry Avenue, Tucson, AZ 85721}
\altaffiltext{4}{Observatoire Astronomique de Strasbourg, 11 rue de l'Universit\'{e}, 67000 Strasbourg, France}
\altaffiltext{5}{Institute for Astronomy, University of Hawaii at Manoa, Honolulu, HI 96822, USA}
\altaffiltext{6}{Department of Physics, Durham University, South Road, Durham DH1 3LE, England}

\begin{abstract}
	We present a method to infer reddenings and distances to stars, based only on their broad-band photometry, and show how this method can be used to produce a three-dimensional dust map of the Galaxy. Our method samples from the full probability density function of distance, reddening and stellar type for individual stars, as well as the full uncertainty in reddening as a function of distance in the 3D dust map. We incorporate prior knowledge of the distribution of stars in the Galaxy and the detection limits of the survey. For stars in the Pan-STARRS 1 (PS1) $3 \pi$ survey, we demonstrate that our reddening estimates are unbiased, and accurate to $\sim 0.13 \, \mathrm{mag}$ in $\EBV$ for the typical star. Based on comparisons with mock catalogs, we expect distances for main-sequence stars to be constrained to within $\sim$20\% - 60\%, although this range can vary, depending on the reddening of the star, the precise stellar type and its position on the sky. A further paper will present a 3D map of dust over the three quarters of the sky surveyed by PS1. Both the individual stellar inferences and the 3D dust map will enable a wealth of Galactic science in the plane. The method we present is not limited to the passbands of the PS1 survey, but may be extended to incorporate photometry from other surveys, such as 2MASS, SDSS (where available), and in the future, LSST and Gaia.
\end{abstract}

\keywords{dust --- ISM: structure --- stars: distances --- Galaxy: structure --- methods: statistical}

\section{Introduction}
\label{sec:introduction}

A long-standing goal of astronomy has been to understand the structure and formation of galaxies. Studies of external galaxies have begun to paint a detailed, global picture of the forces at work in shaping galaxies, but lack the resolution and sensitivity to probe individual stars. In the Milky Way, meanwhile, measurements of the positions and types of millions of stars have been assembled, though these stars probe only a fraction of our Galaxy's volume. Moreover, the positions of these stars are especially uncertain in the Galactic disk, where the bulk of the stars reside, owing to the presence of dust, which obscures and reddens the light from these stars.

Accordingly, wide-field surveys like the Sloan Digital Sky Survey \citep[SDSS;][]{York2000}, which observed millions of stars, have focused on the structure of the stars at high Galactic latitudes, mostly outside the Galactic disk \citep[e.g.,][]{Juric2008, Ivezic2008}. These studies have revealed an abundance of substructure in the Galactic halo \citep{Belokurov2006}, and have led to new constraints on the structure of the Galaxy's halo \citep[e.g.,][]{Law2010} and the identification of challenges to the standard picture of Galaxy formation \citep[e.g., the Missing Satellites Problem,][]{Simon2007}. Still, the bulk of the Galaxy's stars reside and formed in the disk, and it is unclear how much the Galaxy's halo can inform the processes at work there. \citet{Bovy2012, Bovy2012a, Bovy2012b} derive smooth models for the Galactic disk, using sub-populations observed in the Sloan Extension for Galactic Understanding and Exploration \citep[SEGUE;][]{Yanny2009}. Better photometric distance estimates for large, magnitude-limited samples of heavily dust-obscured stars will aid investigation into the smooth structure of the disk, as well as possible disk substructure.

This paper presents a technique to simultaneously infer the distances and reddenings to stars embedded in dust, to enable study of the properties and structure of our Galaxy's disk from optical surveys of resolved stars. We exploit prior knowledge of the types and distribution of stars in the Galaxy in a Bayesian framework to deliver the full probability density function of distance and reddening and stellar type for each star. We derive a principled Bayesian technique to infer reddening as a function of distance, using stars as tracers of the dust column.

We are not the first in this area. \citet{Marshall2006a} produced a 3D extinction map of the Galactic plane by comparing the 2MASS $J - K_{s}$ stellar colors to those of simulated catalogs based on the Besan\c{c}on model of the Galaxy \citep{Robin2003}. Our work is most similar to that of \citet{Sale2012}, who presents related techniques and applies them to simulated IPHAS data \citep{Drew2005}. The method presented here for obtaining stellar reddenings and distances is also related to that of \citet{Berry2011}, who highlight the large amount of 3D structure in the Galaxy's dust using data from SDSS. The work of \citet{Bailer-Jones2011}, likewise, presents a similar technique using broad-band photometry and Hipparcos parallaxes \citep{Perryman1997}. \citet{Vergely2001} and \citet{Lallement2003} map out the 3D distribution of clouds in the Local Bubble by measuring absorption lines imprinted by the interstellar medium on the spectra of stars with Hipparcos parallaxes. \citet{Lallement2013} uses $\sim 23,000$ stellar parallaxes and reddening estimates from a number of sources to infer the 3D distribution of dust opacity out to $800$ - $1000 \, \mathrm{pc}$ in the plane of the Galaxy, and $\sim 300 \, \mathrm{pc}$ out of the plane. Our work differentiates itself from these primarily in that it is adapted to studying the approximately one billion stars with high-quality Pan-STARRS 1 (PS1) photometry, which cover three quarters of the sky and two thirds of the Galactic plane, representing an unprecedented resource for studies of the Galaxy's disk.

The paper is organized as follows. Section \S \ref{sec:PS1} describes the Pan-STARRS 1 survey. Section \S \ref{sec:los-formalism} develops the Bayesian formalism required to produce a three-dimensional reddening map. Sections \S \ref{sec:likelihoods} and \S \ref{sec:priors} describe the stellar and Galactic models we employ. Section \S \ref{sec:sampling} describes the practical implementation of our model to produce a 3D map. Then, in section \S \ref{sec:tests}, we conduct tests of our method with mock photometry, and in section \S \ref{sec:comparison-with-data}, we validate our model with real photometry.

\section{Pan-STARRS 1 Survey}
\label{sec:PS1}

In this work we derive the distances and reddenings to stars observed by Pan-STARRS 1.  Pan-STARRS 1 is a 1.8-meter optical and near-infrared telescope located on Mount Haleakala, Hawaii \citep{PS1_system, PS1_optics}. The telescope is equipped with the GigaPixel Camera 1 (GPC1), consisting of an array of 60 CCD detectors, each 4800 pixels on a side \citep{PS1_GPCa, PS1_GPCb}. The majority of the observing time is dedicated to a multi-epoch $3 \pi$ steradian survey of the sky north of $\delta = -30^{\circ}$ (Chambers in prep.). The $3 \pi$ survey observes in five passbands $g_{\mathrm{P1}}$, $r_{\mathrm{P1}}$, $i_{\mathrm{P1}}$, $z_{\mathrm{P1}}$, and $y_{\mathrm{P1}}$, which together span 400--1000~nm \citep{PS_lasercal}. The images are processed by the Pan-STARRS 1 Image Processing Pipeline (IPP) \citep{PS1_IPP}, which performs automatic astrometry \citep{PS1_astrometry} and photometry \citep{PS1_photometry}.  The data is photometrically calibrated to better than 1\% accuracy \citep{Schlafly2012, JTphoto}.  The resulting homogeneous optical and near-infrared coverage of three quarters of the sky makes the Pan-STARRS1 data ideal for studies of the distribution of the Galaxy's dust.

\section{Line-of-Sight reddening profile}
\label{sec:los-formalism}

Here, we describe the basic assumptions that we make in order to produce a three dimensional dust map. We show that the problem can be decomposed into two steps. In the first step, we determine the probability density function of distance and reddening for each star (See \S \ref{sec:indiv-stars}). In the second step, we use information from stars on the same small patch of sky to infer reddening as a function of distance in the given direction.

We make the assumption that stars which are close to one another in 3D space are behind the same column of dust. By grouping together stars which are close-by on the sky, we are able to use the stars as tracers of the total dust column at different distances in a particular direction on the sky. So long as the dust column does not vary significantly over small angular scales, then this is a valid assumption.

Let us denote the reddening profile along a particular line of sight by
\begin{align}
	E \arg{\mu \, ; \, \vec{\alpha}} \, ,
\end{align}
where $E$, the color excess in some pair of passbands (e.g., $\EBV$), is a function of distance modulus $\mu$, and $\vec{\alpha}$ denotes any fitting parameters defining the reddening profile. These parameters could be, for example, the dust density in each distance bin, and could in principle include the value of $R_{V}$, which parameterizes the wavelength dependence of the extinction law \citep{Cardelli1989, Fitzpatrick1999}. The extinction $\vec{A}$ in any set of passbands is then assumed to be a function of the reddening $E$:
\begin{align}
	\vec{A} = \vec{A} \arg{E , R_{V}} \, .
\end{align}

Along one line of sight, denote the photometry of star $i$ by $\vec{m}_{i}$, and the set of all observed stellar magnitudes as $\left\{ \vec{m} \right\}$. We wish to determine how the model parameters $\vec{\alpha}$ for the line-of-sight reddening profile depend on the stellar photometry $\left\{ \vec{m} \right\}$. That is, we wish to determine
\begin{align}
	p \arg{\vec{\alpha} \, | \left\{ \vec{m} \right\}} \, .
\end{align}
Using Bayes' rule,
\begin{align}
	p \arg{\vec{\alpha} \, | \left\{ \vec{m} \right\}}
	&= \frac{ p \arg{\left\{ \vec{m} \right\} | \, \vec{\alpha}} p \arg{\vec{\alpha}} }{ p \arg{\left\{ \vec{m} \right\}} } \, .
	\label{eqn:alpha-posterior-Bayes}
\end{align}
The likelihood $p \arg{\left\{ \vec{m} \right\} | \, \vec{\alpha}}$ is the probability density of obtaining the set of observed magnitudes $\left\{ \vec{m} \right\}$, given the reddening profile defined by $\vec{\alpha}$. The likelihood of the entire set of stellar observations is just the product of the individual likelihoods, i.e.
\begin{align}
	p \arg{\left\{ \vec{m} \right\} | \, \vec{\alpha}}
	&= \prod_{i} p \arg{\vec{m}_{i} | \, \vec{\alpha}} \, .
\end{align}
This is just a statement that the photometry of one star does not influence the photometry of any other star. Plugging this into Eq. \eqref{eqn:alpha-posterior-Bayes}, and dropping the normalizing factor $p \arg{\left\{ \vec{m} \right\}}$,
\begin{align}
	p \arg{\vec{\alpha} \, | \left\{ \vec{m} \right\}}
	&\propto p \arg{\vec{\alpha}} \, \prod_{i} p \arg{\vec{m}_{i} | \, \vec{\alpha}} \, .
	\label{eqn:alpha-posterior-separated}
\end{align}

We now introduce nuisance parameters describing the distance to and intrinsic type of each star, and then marginalize over these parameters to obtain the likelihood. We denote the distance modulus to star $i$ by $\mu_{i}$, and the parameters describing the stellar type by $\vec{\Theta}_{i}$. For an individual star,
\begin{align}
	p \arg{\vec{m} | \, \vec{\alpha}}
	&= \int \! \d \mu \, \d \vec{\Theta} \,
	p \arg{\vec{m} , \mu , \vec{\Theta} \, | \, \vec{\alpha}} \\
	&= \int \! \d \mu \, \d \vec{\Theta} \,
	p \arg{\vec{m} | \, \mu , \vec{\Theta} , \vec{\alpha}} p \arg{\mu , \vec{\Theta} | \, \vec{\alpha}} \\
	&= \int \! \d \mu \, \d \vec{\Theta} \,
	p \arg{\vec{m} | \, \mu , \vec{\Theta} , E \arg{\mu ; \vec{\alpha}}} p \arg{\mu , \vec{\Theta}} \, .
\end{align}
In the last step, we have assumed that the joint prior on the distance and intrinsic stellar type are independent of the reddening profile. Up to a normalizing constant, the above integrand is equivalent to the posterior density
\begin{align}
	p \arg{\mu , E , \vec{\Theta} \, | \, \vec{m}}
\end{align}
for an individual star, where the prior on $E$ is flat. Intuitively, this is because the prior on reddening is on the fitting parameters $\vec{\alpha}$, rather than the reddening of an individual star. After marginalizing over the stellar type $\Theta$,
\begin{align}
	p \arg{\vec{m} | \, \vec{\alpha}}
	&\propto \int \! \d \mu \,\, p \arg{\mu , E \arg{\mu_{i} ; \vec{\alpha}} | \, \vec{m}} \, .
\end{align}
Plugging the above into Eq. \eqref{eqn:alpha-posterior-separated}, we find that the full posterior density for $\vec{\alpha}$ is given by
\begin{align}
	p \arg{\vec{\alpha} \, | \left\{ \vec{m} \right\}}
	&\propto p \arg{\vec{\alpha}} \, \prod_{i} \int \! \d \mu_{i} \,\, p \arg{\mu_{i} , E \arg{\mu_{i} ; \vec{\alpha}} | \, \vec{m}_i} \, , \label{eqn:line-integral-product}
\end{align}
where we have defined the function
\begin{align}
	p \arg{\mu_{i} , E_{i} | \, \vec{m}_i}
	&\equiv \frac{1}{Z_{i}} \! \int \! \d \vec{\Theta}_{i} \,
	p \arg{\vec{m}_i \, | \, \mu_{i} , \vec{\Theta}_{i} , E_{i}} p \arg{\mu_{i} , \vec{\Theta}_{i}} \, ,
	\label{eqn:single-star-posterior}
\end{align}
which is equivalent to the posterior probability density of finding a single star at distance $\mu_{i}$ and reddening $E_{i}$, where the prior on reddening is flat. Here, $Z_{i}$ is a normalizing constant. Effectively, it is the Bayesian evidence for star $i$, a measure of how likely the star is to be drawn from the model. A higher evidence indicates that the data is more consistent with the model, while a low value for $Z_{i}$ indicates that the model does not well describe the object $i$. Our strategy for sampling from a posterior of the form given by Eq. \eqref{eqn:line-integral-product} is to pre-compute Eq. \eqref{eqn:single-star-posterior} for each star by Markov-Chain Monte Carlo (MCMC) sampling, and then to sample from $p \arg{\vec{\alpha} \, | \left\{ \vec{m} \right\}}$. This approach has two advantages. First, it factorizes the full problem into a series of smaller problems of lower dimension, potentially speeding up the computation. The second advantage of this two-step approach is that it allows outlier rejection on the basis of the Bayesian evidence for each star before proceeding to the second step. Point sources which do not fit the chosen stellar model (e.g., blue stragglers, white dwarfs and galaxies mistakenly classified as stars) can be filtered out by imposing a cut on the evidence $Z_{i}$ (See \S \ref{sec:evidence-outliers}).  More principled approaches to reducing the influence of outliers exist \citep[e.g.,][]{Hogg2010a}, though in the context of our problem they are significantly more computationally expensive to implement.

A more general limitation to the approach taken here is that it does not allow the simultaneous fitting of parameters describing the stars and the spatial variation in dust properties. One could imagine simultaneously constraining stellar types and distances, as well as the dust density and $R_{V}$ parameter throughout space. By fitting the dust properties in many voxels simultaneously, one would be able to place priors on the density power spectrum of the dust, and to infer $R_{V}$ as a function of position in the Galaxy. By fitting dust properties throughout the entire volume of the Galaxy simultaneously, one could even attempt to constrain global parameters, such as the dust scale height and scale length. This would represent a hierarchical approach to the problem of creating a 3D dust map \citep[see, for example,][for a discussion of hierarchical Bayesian models]{kru10}. At the highest level in the hierarchy, one has global parameters, which describe the overall dust distribution and density power spectrum. One level below in the hierarchy, one would have parameters describing the dust properties in each voxel in the Galaxy. At the lowest level, one could have the type and distance for each star. Such a hierarchical approach is appealing because it takes into account spatial correlations in dust properties, and because it directly fits the global structure of the Galaxy's dust component. However, this hierarchical approach potentially requires much greater computational power than the approach we take in this paper, as it does not allow one to process each star individually and treat each line of sight separately, greatly increasing the dimensionality of parameter space. This paper therefore confines itself to fitting each star independently, and then combining the information from each star along any given line of sight to determine the reddening profile as a function of distance.

\section{Individual Stars}
\label{sec:indiv-stars}

Now that we have factorized the problem of determining the line-of-sight reddening profile into one of determining $p \arg{\mu_{i} , E_{i} , \vec{\Theta}_{i} \, | \, \vec{m}_i}$ for each star, we need to determine the individual stellar likelihoods and priors. In the following, we will drop the subscript $i$, as it is assumed that we are dealing with one star.

Using Bayes' Rule,
\begin{align}
	p \arg{\mu , E , \vec{\Theta} \, | \, \vec{m}}
	&\propto p \arg{\vec{m} \, | \, \mu , E , \vec{\Theta}} p \arg{\mu , E , \vec{\Theta}}
\end{align}
The likelihood, $p \arg{\vec{m} \, | \, \mu , E , \vec{\Theta}}$, is the probability density of a star having apparent magnitudes $\vec{m}$, given a distance, reddening and stellar type. The likelihood is thus dependent on our model of intrinsic stellar colors, which we discuss below (in \S \ref{sec:likelihoods}).
The priors, $p \arg{\mu , E , \vec{\Theta}}$, are dependent on our model of the distribution of stars of different types throughout the Galaxy. We discuss our Galactic model in \S \ref{sec:priors}.

\subsection{Stellar Model}
\label{sec:likelihoods}

In our model, each star is described by two intrinsic parameters, its absolute magnitude $M_{r}$ in the PS1 $r$ band and its metallicity $\FeH$. In terms of our previous notation, $\vec{\Theta} = \left( M_{r}, \FeH \right)$. Given a set of stellar templates, $\vec{M} \arg{M, \FeH}$, which map intrinsic stellar type to a set of absolute magnitudes, one obtains theoretical apparent magnitudes
\begin{align}
	\vec{m}_{\mathrm{mod}} = \vec{M} \arg{M_{r} , \FeH} + \vec{A} \arg{E , R_{V}} + \mu \, .
\end{align}
The extinction vector $\vec{A} \arg{E , R_{V}}$ used in this work follows a \citet{Fitzpatrick1999} reddening law with $R_{V} = 3.1$, adapted to the PS1 $grizy_{\mathrm{P1}}$ filter set by \citet{Schlafly2011}. The likelihood of observing apparent magnitudes $\vec{m}$ with Gaussian uncertainties $\vec{\sigma}$ is then given by
\begin{align}
	p \arg{\vec{m} \, | \, \mu , M_{r} , \FeH , E}
	&= \mathcal{N} \arg{\vec{m}  \, | \, \vec{m}_{\mathrm{mod}} \, , \vec{\sigma}} \, .
\end{align}
In this paper, we use the notation $\mathcal{N} \arg{\vec{x} \, | \, \vec{\mu} , \, \vec{\sigma}}$ to denote the probability density of a multivariate normal with mean $\vec{\mu}$ and standard deviations $\vec{\sigma}$, evaluated at $\vec{x}$.

We adopt a set of empirical stellar templates based on stellar observations in PS1, with photometric parallaxes and metallicities derived from the work of \citet{Ivezic2008}. That work determines the absolute magnitude of a star as a function of its intrinsic color and metallicity using observations of globular clusters in SDSS. These globular clusters are uniformly old, and as a result our stellar templates are appropriate only for old populations and do not include age as a parameter. This means that young blue stars are not included, and that the morphology of the subgiant and giant branches is only approximate. Accordingly, our giant branch distances are less reliable than our main sequence distances. Nevertheless, we include the giant branches in our models because any given star may indeed be a distant giant rather than a nearby dwarf.

In detail, we fit a spline to the colors of stars (in 4-color space) near the north Galactic pole to derive the shape of the stellar locus.  All main-sequence stars are required to have intrinsic colors lying along this one-dimensional curve. We then associate each position along the main sequence with an absolute magnitude and a metallicity using the relations of \citet{Ivezic2008}, which give absolute magnitude as a function of color and metallicity. Models for the giants are obtained from linear fits of absolute magnitude to color and metallicity by Ivezi{\'c} (private communication), based on observations of globular clusters. These giant branch fits are joined to the main sequence via a cubic interpolating polynomial for $4 > M_r > 2.35$.  We are able to use relations derived from SDSS because of the close similarity between the PS1 and SDSS filter sets; we transform from the PS1 to SDSS colors using the color transformations of Finkbeiner et al. (in prep.), which have residuals of less than about 1\% across the full range of stellar types considered in this work.

\begin{figure}
	\plotone{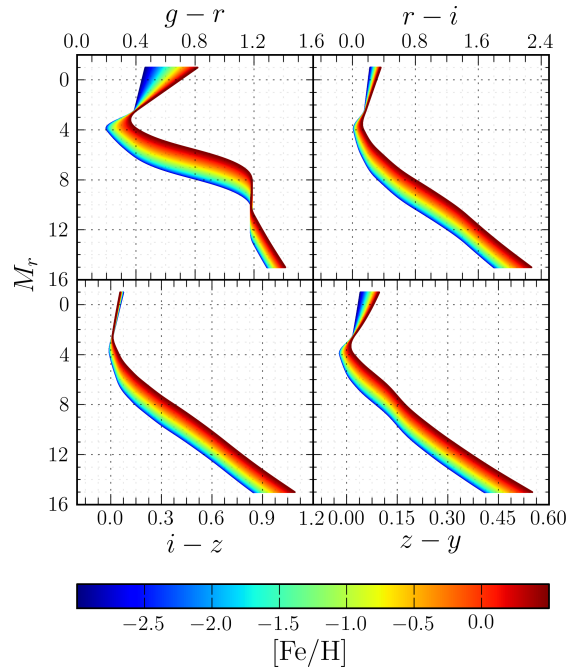}
	\caption{Model stellar colors as a function of absolute $r$-magnitude and metallicity in Pan-STARRS 1 passbands. The stellar templates are based on PS1 color-color relations, and color is related to absolute magnitude and metallicity by SDSS observations of globular clusters \citep{Ivezic2008}. Our empirical templates therefore assume an old stellar population. While the main sequence below the turnoff is nearly invariant with age, the giant branch and the location of the turnoff do, in reality, vary considerably with age. For this reason, we expect our inferences for main-sequence stars to be more accurate than those for giants. The narrowness of the kink at $M_r \simeq 2.4$ is an artifact of our models (See \S \ref{sec:likelihoods}). \label{fig:Mr-vs-color}}
\end{figure}

The resulting stellar templates are shown in Figure~\ref{fig:Mr-vs-color}, which gives the templates' colors as a function of their absolute magnitude and metallicity. Our empirical approach produces a close match to the observed colors of stars, and comparison with Globular and Open Clusters indicates that the absolute magnitudes are accurate along the main sequence (See \S \ref{sec:distance-comparison}). An alternative approach would have been to adopt template colors from a library of synthetic spectra, such as the Padova \& Trieste Stellar Evolution Code \citep[PARSEC;][]{Bressan2012}, giving us access to age as an additional stellar parameter. However, the synthetic libraries have difficulty reproducing the colors of M-dwarfs in detail. We choose therefore to adopt a set of empirical models that match the colors of most stars well, though in future work a hybrid approach may be best suited to the problem.

\subsection{Galactic Model}
\label{sec:priors}

We now present the priors that we place on the intrinsic and extrinsic parameters describing each star. We factorize the priors as follows:
\begin{align}
	p \arg{\mu , M_{r} , \FeH} &= p \arg{\mu} p \arg{\FeH | \, \mu} p \arg{M_{r}} \, .
\end{align}
We describe the distance prior in \S \ref{sec:dist-prior}, the metallicity prior in \S \ref{sec:metallicity-prior} and the prior on absolute magnitude in \S \ref{sec:luminosity-prior}.

\subsubsection{Distance}
\label{sec:dist-prior}

For a given line of sight, the prior probability of finding a star in a small range $\d \mu$ in distance modulus is proportional to the number of stars per unit distance modulus per unit solid angle in the direction of the pixel:
\begin{align}
	p \arg{\mu} \propto \frac{\d N}{\d \mu \d \Omega} = \frac{\d N}{\d r \d \Omega} \frac{\d r}{\d \mu}
	&= n \arg{\mu} r^{2} \, \frac{\d r}{\d \mu} \notag \\
	&\propto 10^{\nicefrac{3 \mu}{5}} \, n \arg{\mu} \, . \label{eqn:p-mu-definition}
\end{align}
Here, $r$ denotes physical distance from the Sun, and $n \arg{\mu}$ is the stellar number density at distance modulus $\mu$ along the chosen line of sight. The distance prior is thus controlled by the line-of-sight number density of stars, as well as a volume factor, $10^{\nicefrac{3 \mu}{5}}$, which grows with distance. This latter term takes into account that the volume represented by a beam of constant, small width in distance modulus grows with distance. For a typical line of sight, the prior is driven upwards by the volume factor at small distances, while farther out it is suppressed by the decline in density in the outer reaches of the Galaxy. For constant density, $n \arg{\mu}$, Eq. \eqref{eqn:p-mu-definition} simply reduces to the Euclidean counts equation.

The distance prior thus requires us to calculate the stellar number density at arbitrary locations in the Galaxy. We employ the three-component Galactic model developed in \citet{Juric2008}, which comprises a thin disk, a thick disk and an oblate halo. In cylindrical coordinates centered on the Galactic center, and with the Galactic plane defining $Z = 0$, each disk component has number density of the form
\begin{align}
	n_{i} \arg{R , Z} &= n_{0 , i} \, e^{-\left( \nicefrac{R}{R_{0 , i}} + \nicefrac{\left| Z \right|}{Z_{0 , i}} \right)} \, ,
	\label{eqn:exp-disk}
\end{align}
where $R_{0 , \, i}$ and $Z_{0 , \, i}$ are the scale radius and height respectively, of each disk component, and $n_{0 , i}$ is the stellar number density of each component at the Galactic center. In the Solar neighborhood,
\begin{align}
	n_{i} \arg{R_{\odot} , \, Z_{\odot}} &= n_{0 , i} \, e^{-\left( \nicefrac{R_{\odot}}{R_{0 , i}} + \nicefrac{\left| Z_{\odot} \right|}{Z_{0 , i}} \right)} \, .
\end{align}
We can thus write the number density of each disk component in terms of locally defined quantities, which are more readily measurable than quantities defined at the Galactic center:
\begin{align}
	n_{i} \arg{R , Z} &= n_{i} \arg{R_{\odot} , \, Z_{\odot}} \, e^{-\left[ \nicefrac{\left( R - R_{\odot} \right)}{R_{0 , i}} + \nicefrac{\left( \left| Z \right| - \left| Z_{\odot} \right| \right)}{Z_{0 , i}} \right]} \, .
\end{align}
where $R_{\odot}$ and $Z_{\odot}$ are the Galactocentric Solar radius and height in cylindrical coordinates, respectively. From this point onwards, we will denote $n_{\mathrm{thin}} \arg{R_{\odot} , \, Z_{\odot}}$ as $n_{\odot}$, and write
\begin{align}
	n_{\mathrm{thick}} \arg{R_{\odot} , \, Z_{\odot}} \equiv f_{\mathrm{thick}} \, n_{\odot} \, .
\end{align}
The halo is assumed to have stellar number density
\begin{align}
	n_{\mathrm{halo}} \arg{R , Z} &= n_{\odot} \, f_{h} \left( \frac{R_{\mathrm{eff}}}{R_{\odot}} \right)^{\!\! -\eta} \, ,
\end{align}
with
\begin{align}
	R_{\mathrm{eff}} &\equiv \sqrt{R^{2} + \left( \nicefrac{Z}{q_{h}} \right)^{2} + R_{\epsilon}^{2}} \, .
\end{align}
Here, $q_{h}$ controls the oblateness of the halo and $\eta$ controls the steepness of the power law. Following \citet{Sesar2011}, the power law breaks at $R_{\mathrm{eff}} = R_{\mathrm{br}}$, becoming steeper. We therefore define $\eta_{\mathrm{inner}}$ and $\eta_{\mathrm{outer}}$, corresponding to the halo steepness inside and outside of the break. We introduce the distance scale $R_{\epsilon}$ over which the inner region of the halo is smoothed, in order to remove the singularity at the Galactic Center. We choose $R_{\epsilon}$ to be $500 \, \mathrm{pc}$; at this scale, it has a negligible effect on the halo density in the regions which \citet{Juric2008} studied, but it prevents the halo from dominating over the disk at the Galactic Center and is the same scale adopted by \citet{Robin2003}.

\begin{figure}
	\plotone{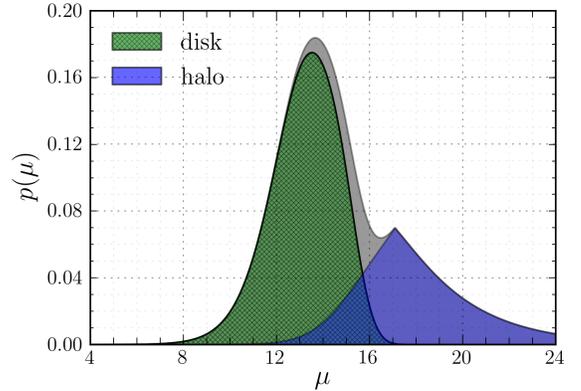}
	\caption{The distance prior for $\left( \ell , \, b \right) = \left( 90^{\circ} , \, 10^{\circ} \right)$. The contributions of the disk and halo are shown individually in green and purple, respectively, while the total prior is given by the gray contour. The break in the contribution from the halo is due to the use of a broken power law for the number density of stars in this component. \label{fig:mu-prior}}
\end{figure}

We employ the parameters given in Table 10 of \citet{Juric2008} and \citet{Sesar2011}. These are listed in Table \ref{tab:density-params}. The adopted value of $f_{h}$, on the low end of the possible range inferred in \citet{Juric2008}, is chosen to better match observed PS1 color-magnitude diagrams at high Galactic latitudes. Throughout, we use $R_{\odot} = 8 \kpc$ and $Z_{\odot} = 25 \pc$. The shape of the distance prior for a line of sight centered on $\ell = 90^{\circ}$ and $b = 10^{\circ}$ is shown in Fig. \ref{fig:mu-prior}.

\begin{deluxetable}{c l || c l || c l}
	\tablewidth{0pt}
	\tablecolumns{8}
	\tabletypesize{\scriptsize}
	\tablecaption{Stellar Number Density Parameters \label{tab:density-params}}
	\tablehead{\multicolumn{2}{c ||}{Thin Disk} & \multicolumn{2}{c ||}{Thick Disk} & \multicolumn{2}{c}{Halo}}
	\startdata
	$R_{\mathrm{thin}}$			&	$2150 \pc$	&
	$R_{\mathrm{thick}}$		&	$3261 \pc$	&
	$R_{\mathrm{br}}$\tablenotemark{$\dagger$}	&	$27.8 \, \mathrm{kpc}$	\\
	$Z_{\mathrm{thin}}$			&	$245 \pc$	&
	$Z_{\mathrm{thick}}$		&	$743 \pc$	&
	$q_{h}$\tablenotemark{$\dagger$}	&	$0.70$		\\
								&				&
	$f_{\mathrm{thick}}$		&	$0.13$		&
	$f_{h}$						&	$0.003$	\\
								&				&
								&				&
	$\eta_{\mathrm{inner}}$\tablenotemark{$\dagger$}	&	$2.62$	\\
								&				&
								&				&
	$\eta_{\mathrm{outer}}$\tablenotemark{$\dagger$}	&	$3.80$
	\enddata
	\tablenotetext{$\dagger$}{Values from \citet{Sesar2011}. All other adopted values are from \citet{Juric2008}.}
\end{deluxetable}

\subsubsection{Metallicity}
\label{sec:metallicity-prior}

We adopt the model of Galactic metallicity developed in \citet{Ivezic2008} and \citet{Bond2010}, assigning separate metallicity distributions to the disk and halo. The metallicity distribution of the disk varies with height above the Galactic plane, and is thus dependent on line-of-sight distance. The metallicity prior takes the form
\begin{align}
	p \arg{\FeH | \, \mu} &= p \arg{\FeH | \, \mu , \, \mathrm{disk}} p \arg{\mathrm{disk} \, | \, \mu} \notag \\
	&\hspace{1cm} + p \arg{\FeH | \, \mathrm{halo}} p \arg{\mathrm{halo} \, | \, \mu} \, .
\end{align}
The membership probabilities are simply
\begin{align}
	p \arg{\mathrm{disk} \, | \, \mu} &= \frac{n_{\mathrm{thin}} \arg{\mu} + n_{\mathrm{thick}} \arg{\mu}}{n_{\mathrm{thin}} \arg{\mu} + n_{\mathrm{thick}} \arg{\mu} + n_{\mathrm{halo}} \arg{\mu}} \, , \\
	p \arg{\mathrm{halo} \, | \, \mu} &= 1 - p \arg{\mathrm{disk} \, | \, \mu} \, ,
\end{align}
which can be calculated on the basis of the preceding discussion (\S \ref{sec:dist-prior}).

\begin{figure}
\plotone{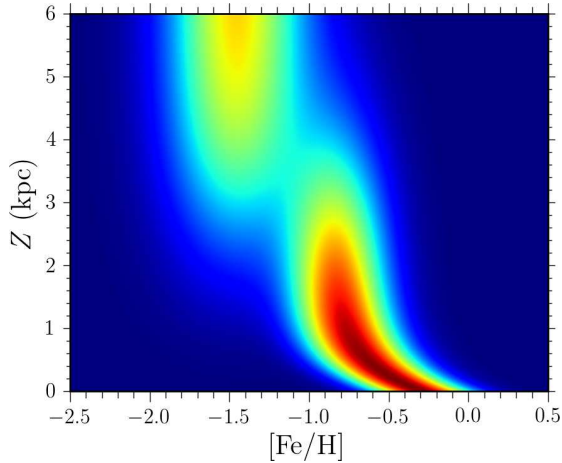}
\caption{The metallicity prior, $p \arg{\FeH | \, Z}$, in the Solar neighborhood ($R = 8 \kpc$). High above the plane of the Galaxy, where the halo dominates, the metallicity distribution has a constant mean and variance. In the plane, where the disk dominates, the mean decreases with scale height. Adapted from Fig. 9 of \citet{Ivezic2008}. \label{fig:FeH-prior}}
\end{figure}

For the disk, stellar metallicity is distributed as a sum of Gaussians. The mean of each Gaussian varies with height above the Galactic plane, so that the prior is best written in terms of cylindrical coordinates:
\begin{align}
	p \arg{\FeH | \, Z , \, \mathrm{disk}} &= c \ \mathcal{N} \arg{\FeH | \, a \arg{Z} \! , \sigma_{D}} \notag \\
	& \hspace{-0.5cm} + (1 \! - \! c) \ \mathcal{N} \arg{\FeH | \, a \arg{Z} \! + \! \Delta a , \sigma_{D}} ,
\end{align}
where
\begin{align}
	a \arg{Z} &= a_{D} + \Delta_{\mu} e^{-\nicefrac{\left| Z \right|}{H_{\mu}}} \, .
\end{align}
The parameters $a_{D}$, controlling the central disk metallicity, and $\Delta_{\mu}$ and $H_{\mu}$, describing the vertical metallicity gradient in the disk, are defined in Table \ref{tab:metallicity-params}. We assume the halo metallicity to be spatially invariant, and distributed as
\begin{align}
	p \arg{\FeH | \, \mathrm{halo}} = \mathcal{N} \arg{\FeH | \, a_{H} , \sigma_{H}} \, .
\end{align}
The parameters, $a_{H}$ and $\sigma_{H}$, describing the halo metallicity, are given in Table \ref{tab:metallicity-params}. The metallicity is plotted in Fig. \ref{fig:FeH-prior} as a function of height above the Galactic midplane in the Solar neighborhood.

\begin{deluxetable}{c r || c r}
	\tablecolumns{4}
	\tabletypesize{\scriptsize}
	\tablecaption{Metallicity Parameters \label{tab:metallicity-params}}
	\tablehead{\multicolumn{2}{c ||}{Disk} & \multicolumn{2}{c}{Halo}}
	\startdata
	$a_{D}$			&	$-0.89$		&	$a_{H}$			&	$-1.46$	\\
	$\sigma_{D}$	&	$0.20$		&	$\sigma_{H}$	&	$0.30$	\\
	$c$				&	$0.63$		&					&			\\
	$\Delta a$	&	$0.14$		&					&			\\
	$\Delta_{\mu}$	&	$0.55$		&					&			\\
	$H_{\mu}$		&	$0.5 \kpc$	&					&			
	\enddata
\end{deluxetable}

\subsubsection{Absolute Magnitude}
\label{sec:luminosity-prior}

We use the $r$-band absolute magnitude to parameterize the luminosity of each star. The joint prior on luminosity and distance is
\begin{align}
	p \arg{\mu , \, M_{r}} \propto \frac{\d N \arg{\mu , \, M_{r}}}{\d \mu \, \d M_{r}} \, .
\end{align}
The luminosity function is assumed to be same in the halo and both disk components, and furthermore independent of position. The priors on distance and luminosity are then separable, so that
\begin{align}
	p \arg{\mu , \, M_{r}} &= p \arg{\mu} p \arg{M_{r}} \, ,
\end{align}
with $p \arg{M_{r}} = \mathrm{LF} \arg{M_{r}} \propto \frac{\d N}{\d M_{r}}$.

We adapt the PS1 luminosity functions provided by the Padova \& Triese Stellar Evolution Code \citep[PARSEC]{Bressan2012}, assuming a \citet{Chabrier2001} log-normal initial mass function. We average over luminosity functions for populations with ages of $\tau = 7 \pm 2 \, \mathrm{Gyr}$ and metallicities $\FeH = -0.5 \pm 0.5 \, \mathrm{dex}$. Denote the luminosity function for a population of age $\tau$ and metallicity $\FeH$ as $\mathrm{LF} \arg{M_{r} \, | \, \tau , \, \FeH}$. The luminosity function we adopt is then
\begin{align}
	\mathrm{LF} \arg{M_{r}}
	&\propto \int \! \d \tau \! \int \! \d \!\FeH \,\,
	\mathrm{LF} \arg{M_{r} \, | \, \tau , \, \FeH} \notag \\
	&\hspace{0.5cm} \times \exp \! \left[ - \frac{\left( \tau - \tau_{0} \right)^{2}}{2 \sigma_{\tau}^{2}} - \frac{\left( \FeH - \FeH_{0} \right)^{2}}{2 \sigma_{\FeH}^{2}} \right] \, ,
	\label{eqn:LF-averaging}
\end{align}
with $\tau_{0} = 7 \, \mathrm{Gyr}$, $\sigma_{\tau} = 2 \, \mathrm{Gyr}$, $\FeH_{0} = -0.5 \, \mathrm{dex}$ and $\sigma_{\FeH} = 0.5 \, \mathrm{dex}$. In principle, it is possible to make the luminosity function depend on metallicity, by not averaging over $\FeH$ in Eq. \eqref{eqn:LF-averaging}. For simplicity, we assume here that the luminosity function is universal.

\subsubsection{Reddening}
\label{sec:reddening-prior}

As indicated in \S \ref{sec:los-formalism}, the manner in which we have factorized the line-of-sight reddening problem requires us to place a flat prior on the color excess, $E$, for each star. The priors on the reddening profile are imposed on the parameters which control the line-of-sight reddening, rather than on individual stellar reddenings. For example, if one divides each line of sight into $N$ distance bins and assigns a different dust density $\rho_{i}$ to each bin, then the reddening prior would take the form $p \arg{\rho_{1} , \, \rho_{2} , \, \dots , \, \rho_{N}}$.

\subsubsection{Survey Selection Function}
\label{sec:malmquist-bias}

The distance prior developed above only asks how many stars are in a thin shell at each distance. However, for a magnitude-limited survey, we would like instead to know the number of observable stars at a given distance. We should assign zero prior probability to the possibility of a star being observed which our instrument cannot detect. The fact that a star has been observed by a given instrument therefore tells us something about its stellar type, distance and extinction. Using the notation from \citet{Sale2012}, we define the vector $\vec{S}$ for each star, where $S_{i}$ is true if a star has been observed in passband $i$, and false if the star is not detected in that passband. The PS1 dataset used in this paper is not based on forced photometry, so there is a separate probability of a source being detected in each passband. If forced photometry were conducted, there would be one single probability $p \arg{S}$, equal to the probability of detecting the source in at least one of the passbands. Including this information in the single-star posterior, Eq. \eqref{eqn:single-star-posterior}, we get
\begin{align}
	p \arg{\mu , E , \vec{\Theta} \, | \, \mobs , \vec{S}}
	&\propto p \arg{\mobs | \, \mu , E , \vec{\Theta} , \vec{S}} \notag \\
	&\hspace{1.3cm} \times p \arg{\mu , E , \vec{\Theta} \, | \, \vec{S}} \, .
\end{align}
But the prior is now just
\begin{align}
	p \arg{\mu , E , \vec{\Theta} \, | \, \vec{S}}
	&\propto
	p \arg{\vec{S} \, | \, \mu , E , \vec{\Theta}} p \arg{\mu , E , \vec{\Theta}} \, ,
\end{align}
so in full,
\begin{align}
	p \arg{\mu , E , \vec{\Theta} \, | \, \mobs , \vec{S}}
	&\propto p \arg{\mobs | \, \mu , E , \vec{\Theta} , \vec{S}} p \arg{\mu , E , \vec{\Theta}} \notag \\
	&\hspace{1.5cm} \times p \arg{\vec{S} \, | \, \mu , E , \vec{\Theta}} \, .
\end{align}
The first term is simply the likelihood we found earlier, since the knowledge that the star has been detected has no effect on the apparent magnitudes the model predicts, assuming the stellar type, distance and reddening are known. That is to say,
\begin{align}
	\vec{m}_{\mathrm{mod}} &= \vec{M}_{\mathrm{mod}} \arg{\vec{\Theta}} + \vec{A} \arg{E} + \mu \, ,
\end{align}
and
\begin{align}
	p \arg{\mobs | \, \mu , E , \vec{\Theta} , \vec{S}} &= \mathcal{N} \arg{\mobs | \, \vec{m}_{\mathrm{mod}} , \vec{\sigma}} \, .
\end{align}
The only element of the calculation which changes when we take into account Malmquist bias is therefore the prior, which picks up an extra factor of
\begin{align}
	p \arg{\vec{S} \, | \, \mu , E , \vec{\Theta}} &= p \arg{\vec{S} \, | \, \vec{m}_{\mathrm{mod}}} \notag \\
	&= \prod_{i} p \, \arg{S_{i} \, | \, m_{\mathrm{mod} , \, i}} \, .
	\label{eqn:merged-det-prob}
\end{align}
This is the survey selection function.

If forced photometry were used instead, then we would have a single detection parameter $S$, denoting that the source was detected in at least one passband, and the survey selection function would be
\begin{align}
	p \arg{S \! = \! \mathrm{true} \, | \, \mu , E , \vec{\Theta}} &= p \arg{S \, | \, \vec{m}_{\mathrm{mod}}} \notag \\
	&= 1 - \prod_{i} p \, \arg{S_{i} \! = \! \mathrm{false} \, | \, m_{\mathrm{mod} , \, i}}
\end{align}
in place of the expression in Eq. \eqref{eqn:merged-det-prob}.

We therefore require an estimate of the completeness of the survey in each band, as a function of apparent magnitude. We determine completeness by comparison with point-source detections in the $275 \, \mathrm{deg}^{2}$ SDSS Stripe 82 survey \citep{York2000, Annis2011}. The co-added Stripe 82 images go more than a magnitude deeper than the individual PS1 $3 \pi$ images, allowing us to use Stripe 82 detections as a complete catalog of point sources past the detection limits of the PS1 $3 \pi$ survey. For each Stripe 82 source, we deteremine whether there is a Pan-STARRS 1 detection within $1^{\prime \prime}$. The completeness fraction of the PS1 $3 \pi$ survey is the percentage of Stripe-82 detections with a PS1 match.

We determine the completeness fraction of the PS1 $3 \pi$ survey as a function of
\begin{align}
	\Delta m \equiv m - m_{\mathrm{lim}} \, ,
\end{align}
where $m$ is the PS1 magnitude, and $m_{\mathrm{lim}}$ is an estimate of the local PS1 $5 \sigma$ magnitude limit, based on the point-spread function of nearby PS1 detections and local sky and read noise. We divide the SDSS Stripe 82 footprint into HEALPix ${\tt nside} = 128$ pixels (with $\sim 27^{\prime}$ scale). In each pixel, we select all Stripe-82 detections classified as stars, and transform their $ugriz$ magnitudes to $grizy_{\mathrm{P1}}$, using color transformations derived by Finkbeiner (in prep.), based on standard-star catalogs. In each pixel, we determine the PS1 limiting $grizy_{\mathrm{P1}}$ magnitudes from the median limiting magnitudes estimated for individual PS1 detections. For each Stripe 82 detection in the pixel, we obtain $\Delta m$ in each band by subtracting the local limiting magnitude from the transformed detection magnitude. In each passband, we bin Stripe 82 detections by $\Delta m$, obtaining an empirical estimate of the completeness in each bin from the number of PS1 matches.

\begin{figure*}
	\epsscale{0.75}
	\plotone{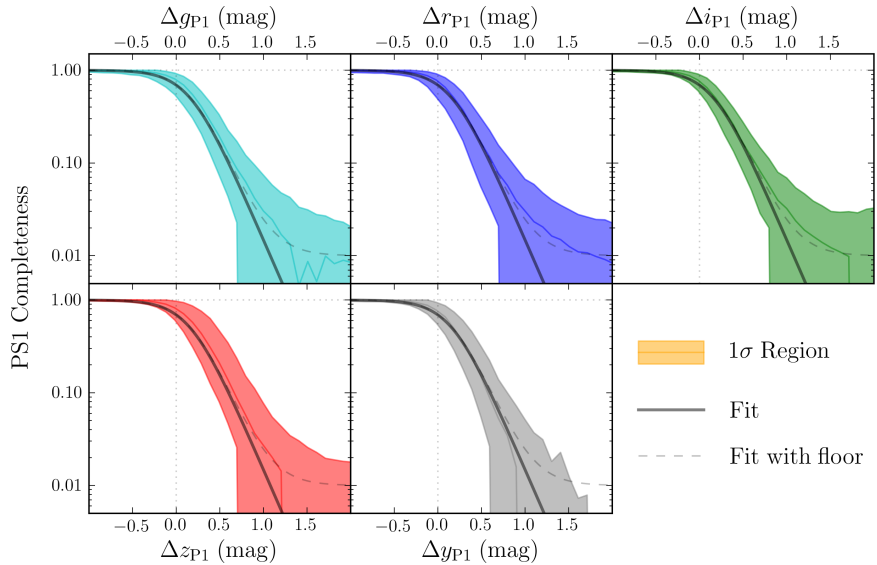}
	\caption{Completeness of the PS1 $3 \pi$ survey, as a function of magnitudes past a locally estimated $5 \sigma$ magnitude limit. The completeness is estimated by comparison with SDSS Stripe 82 \citep{York2000, Annis2011}. The shaded curve shows the median completeness, with $1 \sigma$-range of completeness in each bin, based on estimates in 27' pixels. The solid black line shows our fit to the completeness curve. The dashed black line shows the effect of adding a small floor to our fit, which takes into account an assumed small rate of false coincidences between PS1 and Stripe 82 detections.\label{fig:completeness}}
\end{figure*}

We find that the completeness fraction is reasonably well fit by
\begin{align}
	& p \arg{S_{i} = \mathrm{true} \, | \, m_{\mathrm{mod} , \, i}} \notag \\
	& \hspace{0.5cm} = \left[ 1 + \exp \arg{ \frac{m_{\mathrm{mod} , \, i} - m_{\mathrm{lim} , \, i} - \Delta m_{1}}{\Delta m_{2}} } \right]^{-1} ,
	\label{eqn:completeness-fraction}
\end{align}
where $m_{\mathrm{lim} , \, i}$ is a limiting magnitude calculated for each point-source detection in the PS1 $3 \pi$ survey, equal to the magnitude of a source that would be detected at $5 \sigma$ in one exposure, given the sky and read noise. $\Delta m_{1} = 0.16 \, \mathrm{mag}$ and $\Delta m_{2} = 0.2 \, \mathrm{mag}$ are fitting parameters. The positive value of $\Delta m_{1}$ indicates that the PS1 pipeline goes somewhat deeper than our naive estimate $m_{\mathrm{lim}}$. The same fitting parameter values reproduce the completeness curve in all five PS1 passbands reasonably well, reflecting the consistency of the PS1 optics and pipeline across the entire filter set. The empirically measured completness fraction and our fit are plotted for each passband in Fig. \ref{fig:completeness}.

\section{Sampling Method}
\label{sec:sampling}

\subsection{Individual Stars}

\begin{figure}
	\plotone{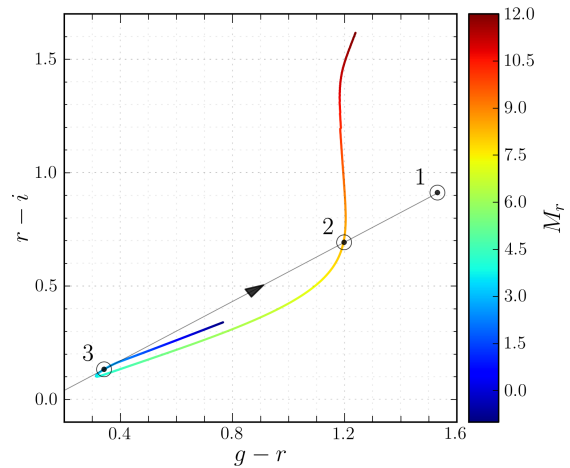}
	\caption{Sketch of how photometric parallax works, for illustrative purposes, adapted from \citet{Berry2011}. A star is observed at location $1$ in color-color space. Its de-reddened colors may lie along any point on the gray line, parallel to the reddening vector. The intersections of this line with the model stellar locus, labeled $2$ and $3$, represent the most likely intrinsic stellar types. The posterior density for the star will thus have two modes -- one at larger distance and lesser reddening ($2$), and one at smaller distance and greater reddening ($3$). For simplicity, we assume Solar metallicity in this example. This is how one would make a distance and reddening determination by eye. Our more rigorous Bayesian method takes into account photometric uncertainties, as well as priors on stellar type and Galactic structure. \label{fig:reddening-vector}}
\end{figure}

We use Markov Chain Monte Carlo sampling to explore the parameter space for individual stars. The sampling must be performed with great care, owing to two features of the distributions $p \arg{ \mu , E }$. First, the distributions are invariably highly elongated, and second they are often multimodal. The elongation stems from the close alignment between the reddening vector and the stellar locus in the PS1 bands, as shown in Fig.~\ref{fig:reddening-vector}. The multimodality has two causes. First, the reddening vector in general intersects the $gri$ stellar locus in two locations, creating a degeneracy between blue and red main-sequence stars. Second, the PS1 bands do not distinguish dwarfs from giants, leading to the possibility that a star can be either a faraway red giant or a nearby red dwarf.

These degeneracies are easier to visualize if we consider only three passbands, as shown in Figure~\ref{fig:reddening-vector}. In this reduced space, stellar photometry is fully described by a single overall observed magnitude and two colors. If we observe a star at a given location in color-color space, we can then move backwards along the reddening vector until we intersect the stellar locus. One can then compare the observed apparent magnitude with the absolute magnitude of the stellar locus at the point of intersection. One thus obtains both a reddening and distance for the star. If there are multiple intersections, then the observed star could be of different intrinsic types, and thus have different reddening and distance combinations. The relative probability of each mode is in practice given by the space of stellar types lying close to the gray line, as well as the priors applied to the problem.

We perform a Markov Chain Monte Carlo sampling of these surfaces using a custom C++ implementation of the affine-invariant sampler introduced in \citet{Goodman2009} and recently given in a python implementation by \citet{Foreman-Mackey2012}. We employ both short-range ``stretch'' and long-range ``replacement'' moves \citep{Goodman2009}. The long-range moves allow mixing between widely separated modes in parameter space, but are more computationally expensive than the short-range ``stretch'' steps. The replacement moves are closely related to the Normal Kernel Coupler of \citet{Warnes2001}. We have found that with the addition of long-range ``replacement'' steps, the affine-invariant sampler is capable of handling the multimodality of the problem, and that it is well suited to the strong degeneracies in parameter space. For each star, we sample from each stellar posterior density in four independent runs and check convergence with the Gelman-Rubin diagnostic \citep{Gelman1992}. The Gelman-Rubin diagnostic essentially verifies that the variance between the means of separate chains is small compared with the variance within the chains. If independent MCMC runs produce significantly different estimates of the parameter means, one or more of the runs must not have converged. Each run employs 20 samplers, with a mix of 80\% stretch steps and 20\% replacement steps, and 2000 steps per sampler. The first 1000 steps from each sampler are discarded as burn-in. Thus, excluding the burn-in phase, a total of 80000 samples are drawn for each star across the four chains, with the number of independent samples being lower. On four cores of a 2.67 GHz Intel Xeon X5650 with 12 MB of L3 cache, our run time per star is typically 0.15 seconds per run per core, or 0.8 CPU seconds for four independent runs.

\subsection{Bayesian Evidence \& Outlier Rejection}
\label{sec:evidence-outliers}

When an observed object does not match our stellar model, the inferences we draw on its distance, reddening and stellar type are unreliable. One means of quantifying the reliability of our inferences for an individual star is to compute the evidence
\begin{align}
	Z \equiv \int \! \d \mu \, \d E \, \d \vec{\Theta} \, p \arg{\vec{m} , \vec{S} \, | \, \mu , E , \vec{\Theta}} \, p \arg{\mu , E , \vec{\Theta}} \, ,
\end{align}
which is the probability density of drawing the observed magnitudes $\vec{m}$ from the stellar model and observing the star in the PS1 survey. A low evidence indicates that the observed point source is a member of a stellar population not included in our model (e.g., a young blue giant or an unresolved binary system with colors that do not match any stellar template), is not a star (e.g., a white dwarf or galaxy), that the errors in the photometry have been underestimated, or that the reddening vector is inaccurate. Here, our approach is similar to \citet{Berry2011}, which identifies objects which do not fit the stellar model by a threshold $\chi^{2}$ statistic. There is no direct analogue for the $\chi^{2}$ statistic in a Bayesian framework, but evidence may serve a similar purpose in model comparison.

Note that as we do not include priors on the extinction to individual stars, but rather on the line-of-sight reddening profile, we do not strictly calculate the evidence. Instead, we calculate the evidence of a model with a prior on $E$ with wide support (i.e. a prior which allows $E$ to take on a wide range of values), such that the prior is nearly constant across all relevant reddennings:
\begin{align}
	p \arg{E} \approx
	                     \begin{cases}
	                     	a			&	0 \leq E \lesssim E_{0} \\
	                     	a f \arg{E}	&	E \gtrsim E_{0}
	                     \end{cases}
	                     \, ,
\end{align}
where $f \arg{E}$ is some integrable function whose precise behavior is unimportant, $a$ is a normalizing constant, and $E_{0}$ is some large extinction. We thus effectively calculate the evidence $Z$ for a model of this form, up to a constant factor $a$, which is the same for every star. This allows for outlier rejection, based on comparisons between the evidence for different stars.

We employ a modified harmonic mean estimate, which is obtained directly from the Markov chain produced in sampling the posterior density, and thus requires little additional computation \citep{Gelfand1994, Robert2009}. This method is presented in more detail in Appendix \ref{sec:app-harmonic-mean}.

\subsection{Line-of-Sight Fitting}
\label{sec:los-fitting}

We choose the HEALPix pixelization scheme \citep{Gorski2005} as our method of dividing the sky into individual lines of sight. Once we have determined $p \arg{\mu , \, E \, | \, \vec{m}}$ for each star in a given HEALPix pixel and rejected stars which fall below the evidence cut, we can apply Eq. \eqref{eqn:line-integral-product} to determine the posterior probability of the parameters $\vec{\alpha}$ describing the reddening profile. We parameterize the reddening profile as a piecewise-linear function in distance modulus, with $\alpha_{i} = \Delta E^{\left( i \right)}$ describing the rise in $r$-band reddening in distance segment $i$. We split up each line of sight into 30 distance segments of equal width in $\mu$, with the closest distance being at $\mu = 4$, corresponding to $63 \pc$, and the furthest distance being at $\mu = 19$, corresponding to $63 \kpc$. It must be cautioned that, in general, our method does not tightly constrain reddening at this latter distance, where PS1 observes very few stars. In addition to requiring that reddening increase monotonically with distance, we apply a wide log-normal prior on the differential reddening in each distance bin, $\left\{ \Delta E^{\left( i \right)} \right\}$.

We use the affine-invariant sampler to draw a representative sample of possible reddening profiles. We sample from the posterior density given by Eq. \eqref{eqn:alpha-posterior-separated}. We can thus produce a three-dimensional reddening map which includes the uncertainty in reddening as a function of distance.

\section{Tests with Mock Photometry}
\label{sec:tests}

The first and most straightforward test of our method is to generate mock photometry for stars of varying stellar type, distance and extinction, and to see how well we can recover those parameters. This is less of a test of the particular stellar model used than a demonstration that photometry alone is capable of sufficiently constraining stellar parameters. We find that our method is capable of accurately recovering both single-star parameters and the line-of-sight reddening profile.

\subsection{Generating Mock Catalogs}
In order to generate a mock photometric catalog for a particular region on the sky, we begin by drawing intrinsic stellar types (metallicities and absolute $r_{\mathrm{P1}}$ magnitudes) and distances from our priors. We assign a reddening to each star, either according to an assumed distance -- reddening relationship, or from a reddening distribution we define, depending on the purpose of the mock catalog. For each star in the catalog, we generate model magnitudes, as described in \S \ref{sec:likelihoods}.

We determine which passbands each star is detected in, according to our probabilistic PS1 completeness model, Eq. \eqref{eqn:completeness-fraction}. In the remainder of the paper, we reject simulated stars which do not have 5-band detections.

We then apply magnitude-dependent Gaussian photometric errors to each simulated star to obtain observed magnitudes. The error we apply to each star in each passband is a function of the model apparent magnitude:

\begin{align}
	\sigma^{2} \arg{m}
	= \sigma_{\mathrm{floor}}^{2} + \sigma_{0}^{2} \exp \! \left[ \frac{2 \left( m - m_{\mathrm{lim}} \right)}{\Delta m_{3}} \right] \, .
	\label{eqn:PS1-error-model}
\end{align}

As before, $m_{\mathrm{lim}}$ is the $5 \sigma$ limiting magnitude in the given passband. We set the error floor to $\sigma_{\mathrm{floor}} = 0.02$. For PS1 passbands, we find that $\sigma_{0} = 0.16$ and $\Delta m_{3} = 0.8$ give a reasonable fit to the photometric uncertainties.

Our final catalog thus contains noisy observed magnitudes of each star, along with the photometric uncertainty in each passband. As a final step, we plug the observed magnitudes back into Eq. \eqref{eqn:PS1-error-model} to obtain a new estimate of the photometric uncertainties for each star. The final catalog that we pass to our pipeline thus reports inexact photometric uncertainties, much as a realistic catalog would. The usefulness of these mock catalogs is that they allow us to generate a large amount of photometry for stars whose `true'' distances and reddenings are known.

\subsection{Single-Star Tests}
\label{sec:mock-single-star}

\begin{figure}
	\plotone{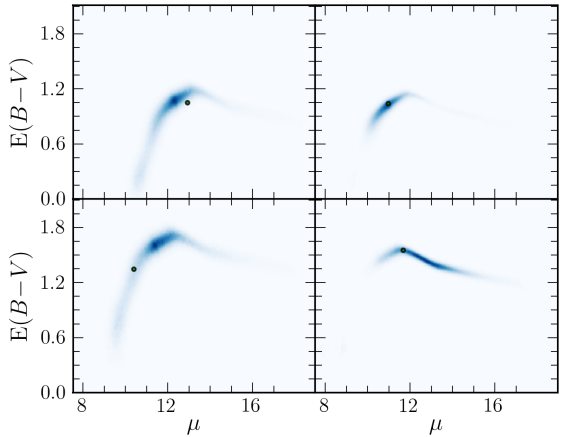}
	\caption{Distance and reddening estimates for four simulated stars. The joint posterior in distance and reddening is shown as a heat map. As this is mock photometry, we know the ``true'' distances and reddenings for the stars, which are shown as green dots. The true stellar parameters lie in regions of high inferred probability, as expected. The shape of the probability density functions traces that of the stellar locus. The probability density at closer distances corresponds to the main sequence, with increasing reddening compensating for the bluer intrinsic colors as one travels up the stellar locus. The peak in reddening corresponds to the main-sequence turnoff. Distances beyond the turnoff correspond to the giant branch. \label{fig:mock-indiv}}
\end{figure}

We illustrate the typical appearance of single-star posterior distributions in distance and reddening in Fig. \ref{fig:mock-indiv}. For each of the four simulated stars in Fig. \ref{fig:mock-indiv}, the ``true'' distance and reddening are indicated by a dot, while the background heat map shows the probability density inferred by our pipeline.

\begin{figure*}
	\plotone{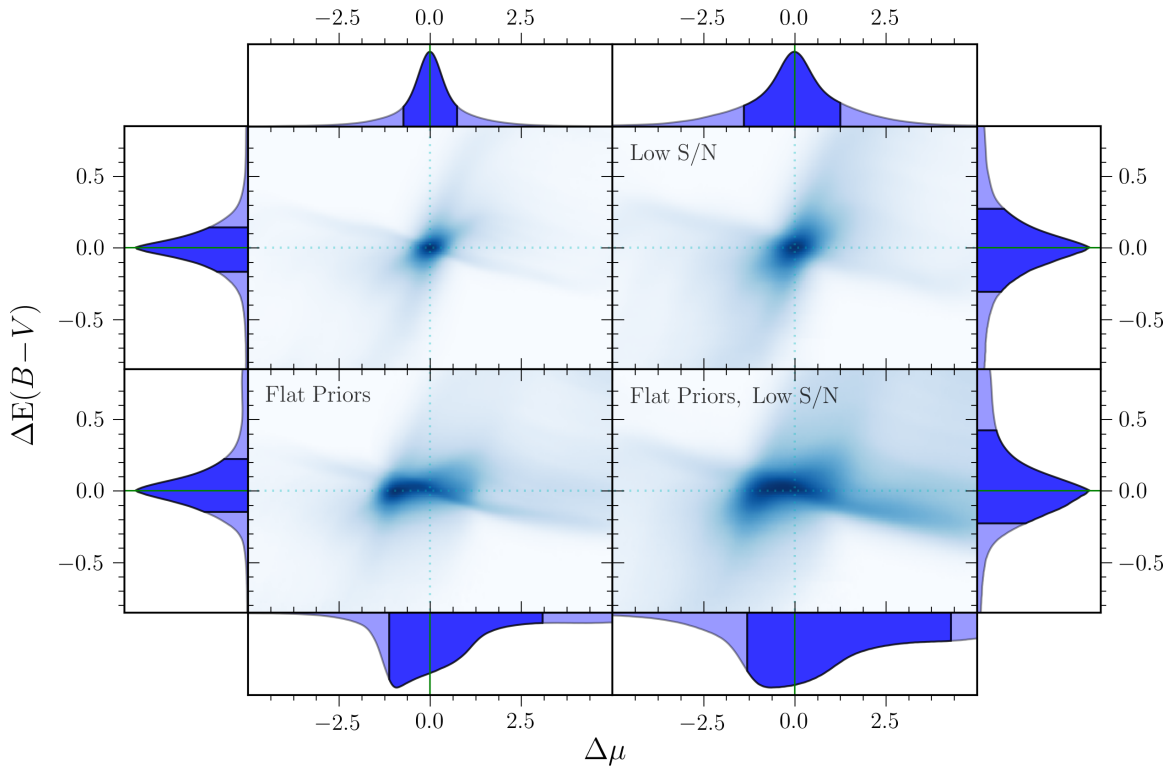}
	\caption{Centered and stacked probability densities on a linear scale for 5000 stars along a simulated line of sight pointed at $\ell = 90^{\circ}$, $b = 20^{\circ}$. In the bottom panels, we fit the stars using flat priors, so that only the likelihood function comes into play. In the right panels, we show inferences for low signal-to-noise detections, generated using $3\times$ the normal observational uncertainties. Removing the priors biases the inferred distances and, to a lesser extent, reddenings. Inferred distances and reddenings for stars with high signal-to-noise detections have smaller uncertainties. In each panel, each stellar probability density function has first been centered on the true distance and reddening, before the probability densities for the stars have been summed, as described in the text. The ``X'' shape of the stacked probability densities in the top-left panel reflects the existence of separate giant and dwarf modes. The feature stretching from the bottom left to the top right corresponds to the main sequence, while the perpendicular feature corresponds to the giant mode. The histograms bordering each panel show the distribution of $\Delta \mu$ and $\Delta \EBV$, with the 15.87\% to 84.13\% region shaded. \label{fig:mock-stack}}
\end{figure*}

In order to determine how far off our estimates are on average, we define the centered probability density
\begin{align}
	\tilde{p} \arg{\Delta \mu , \, \Delta E}
	&\equiv \frac{1}{N} \sum_{i=1}^{N} p \arg{\mu^{\ast}_{i} + \Delta \mu, \, E^{\ast}_{i} + \Delta E} \, ,
\end{align}
where $\mu^{\ast}_{i}$ and $E^{\ast}_{i}$ are the true distance modulus and reddening, respectively, for star $i$. For a simulated line-of-sight, this function gives the average probability density of our inference being offset from the true stellar parameters by $\left( \Delta \mu , \, \Delta E \right)$. We plot $\tilde{p}$ for a typical line of sight in Fig. \ref{fig:mock-stack}. The typical spread of in $\Delta \mu$ and $\Delta E$ varies across different lines of sight, but the centered probability density generally peaks at the origin, as should be expected. In the bottom two panels of \ref{fig:mock-stack}, we show the effect of applying flat priors to the stellar parameters, in place of the priors developed in \S \ref{sec:indiv-stars}. The effect is to widen and bias the inferred probability density functions. Due to the near-alignment of the reddening vector with the PS1 stellar locus for much of the main sequence, this bias remains even for stars with low observational uncertainties. The stellar priors are thus important in correctly inferring stellar parameters from Pan-STARRS 1 photometry. In the right two panels of \ref{fig:mock-stack}, we only use inferences for stars with low signal-to-noise detections. The low signal-to-noise population is generated using an inflated error model that applies $3\times$ the normal observational uncertainty to the mock photometry. As expected, the inferred parameters for such stars are less constrained, but they are nonetheless unbiased.

In Table \ref{tab:mock-comparison}, we present typical uncertainties in the inferred distance and reddening of individual stars. To do this, we generate mock catalogs along two different lines of sight. For our high-Galactic-latitude target, we choose the North Galactic Pole, where the stellar population is dominated by the halo. Here, we apply reddenings of $\EBV \lesssim 0.1$ to the simulated stars. For the low Galactic latitude target, we choose $\ell = 45^{\circ}$, $b = 0^{\circ}$, and draw reddening uniformly from the range $0 \leq \EBV \leq 2$. We run the two mock catalogs through our pipeline, and compare the inferred distances and reddenings, drawn from the posterior probability density $p \arg{\mu , \, A}$, to the true values. For this test, we allow our inferred reddenings to go slightly negative ($\EBV > -0.25$), to avoid introducing a bias into the inferred values. We give the median, and 15.87th and 84.13th percentiles of $\tfrac{\Delta d}{d}$ and $\Delta \EBV$, equivalent to the one-standard-deviation range for a Gaussian distribution.

Uncertainties in distance modulus can be transformed to uncertainties in distance by making use of the relation
\begin{align}
	d = \left( 10 \, \mathrm{pc} \right) 10^{\nicefrac{\mu}{5}} \, .
\end{align}
Let $\mu_{\mathrm{inferred}} = \mu_{\mathrm{true}} + \Delta \mu$. Then,
\begin{align}
	\frac{\Delta d}{d}
	\equiv \frac{d_{\mathrm{inferred}} - d_{\mathrm{true}}}{d_{\mathrm{inferred}}}
	&= 10^{\nicefrac{\Delta \mu}{5}} - 1 \, .
	\label{eqn:Delta-d}
\end{align}
Similarly, we define $\Delta \EBV$ as $\EBV_{\mathrm{inferred}} - \EBV_{\mathrm{true}}$.

Our distance and reddening estimates are unbiased. However, if one selects a subsample of stars of a certain known type, a bias in distance is introduced. Thus, distance estimates for mock dwarf stars are biased low, as the model assigns some probability to the possibility of them being giants. Inversely, distances to giants are biased high. A star drawn at random, however, has an unbiased distance estimate. Distance and reddening constraints depend upon the quality of the photometry and direction on the sky, and therefore vary significantly on a star-per-star basis. It is, in general, more informative to look at the detailed shape of the posterior distribution for a given star in distance and reddening space (See Fig. \ref{fig:mock-indiv}).

\begin{deluxetable}{c c c c c}
	\tablewidth{0pt}
	\tablecolumns{8}
	\tabletypesize{\scriptsize}
	\tablecaption{Uncertainty in Inferred Distances and Reddenings \label{tab:mock-comparison}}
	
	\tablehead{
	\colhead{} &
	\multicolumn{2}{c}{Low Latitude} &
	\multicolumn{2}{c}{High Latitude} \\
	\cline{2-3} \cline{4-5} \\
	\colhead{} &
	\colhead{${\tfrac{\Delta d}{d}}^{\ast}$} & \colhead{$\Delta \EBV$} &
	\colhead{$\tfrac{\Delta d}{d}$} & \colhead{$\Delta \EBV$}
	}
	\startdata
	$ -1 \! < \! M_{r} \! \leq \! 4$ &
	$-20 \substack{+41 \% \\ -30 \%}$ &
	$-0.03 \substack{+0.07 \\ -0.12}$ &
	$-37 \substack{+48 \% \\ -32 \%}$ &
	$-0.03 \substack{+0.09 \\ -0.11}$ \\ \\
	$ 4 \! < \! M_{r} \! \leq \! 6$ &
	$6 \substack{+55 \% \\ -33 \%}$ &
	$0 \substack{+0.12 \\ -0.20}$ &
	$12 \substack{+79 \% \\ -37 \%}$ &
	$0 \substack{+0.10 \\ -0.12}$ \\  \\
	$ 6 \! < \! M_{r} \! \leq \! 8$ &
	$23 \substack{+97 \% \\ -34 \%}$ &
	$0.17 \substack{+0.28 \\ -0.36}$ &
	$8 \substack{+92 \% \\ -21 \%}$ &
	$0.04 \substack{+0.22 \\ -0.20}$ \\ \\
	$ 8 \! < \! M_{r} \! \leq \! 10$ &
	$33 \substack{+261 \% \\ -39 \%}$ &
	$0.29 \substack{+0.62 \\ -0.42}$ &
	$4 \substack{+31 \% \\ -14 \%}$ &
	$0.02 \substack{+0.33 \\ -0.09}$ \\ \\
	$ 10 \! < \! M_{r} \! \leq \! 12$ &
	$5 \substack{+14 \% \\ -15 \%}$ &
	$0.02 \substack{+0.13 \\ -0.13}$ &
	$1 \substack{+15 \% \\ -12 \%}$ &
	$0 \substack{+0.10 \\ -0.09}$ \\ \\
	Dwarfs$^{\dagger}$ &
	$9 \substack{+63 \% \\ -33 \%}$ &
	$0.01 \substack{+0.16 \\ -0.20}$ &
	$4 \substack{+50 \% \\ -18 \%}$ &
	$0.01 \substack{+0.17 \\ -0.12}$ \\ \\
	All Stars &
	$-3 \substack{+55 \% \\ -35 \%}$ &
	$-0.01 \substack{+0.12 \\ -0.16}$ &
	$2 \substack{+45 \% \\ -23 \%}$ &
	$0.01 \substack{+0.16 \\ -0.12}$
	\enddata
	\tablenotetext{$\ast$}{$\tfrac{\Delta d}{d}$ is given in percent. See Eq. \eqref{eqn:Delta-d}.}
	\tablenotetext{$\dagger$}{Dwarfs are defined here as all stars in the range $4 \! < \! M_{r} \! \leq \! 12$.}
\end{deluxetable}

\begin{figure}
	\vspace{0.5cm}
	\epsscale{1.0}
	\plotone{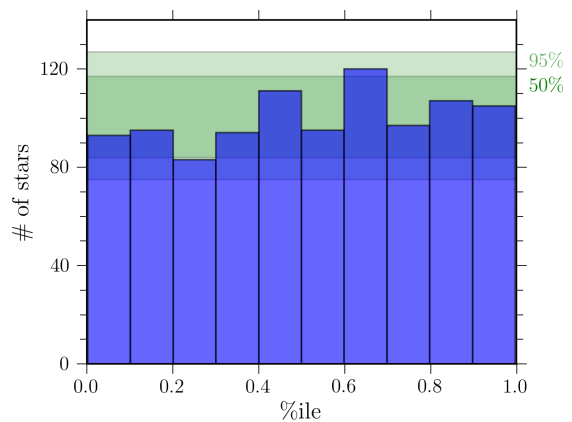}
	\caption{Distribution of percentiles for simulated photometry of 1000 stars, as defined in Eq. \eqref{eqn:pctile-def}. The percentiles are expected to be drawn from the standard uniform distribution, resulting in each bin being of equal height. In 50\% of trials, we would expect all the bins to fall within the dark green band, and in 95\% of trials, all the bins should lie within the light green band. The percentiles are consistent with being drawn from a uniform distribution, indicating that we are sampling from the model correctly. \label{fig:percentiles}}
\end{figure}

Next, we test that the true stellar parameters are drawn from the probability density functions we calculate. For each simulated star, we derive the posterior density $p \arg{\mu , \, E}$, based on the simulated photometry. Since we know the true distance modulus $\mu^{\ast}$ and reddening $A^{\ast}$ of the star, a natural question is whether $\mu^{\ast}$ and $E^{\ast}$ are drawn from $p \arg{\mu , \, E}$. This cannot be answered for a single star, but we can test this hypothesis for a large number of stars. Assign a percentile to a given star as follows:
\begin{align}
	P \arg{p < p^{\ast}} \equiv \int_{p \, \arg{\mu , \, E} < p^{\ast}} \hspace{-1.35cm} \d \mu \, \d E \,\, p \arg{\mu , \, E} \, ,
	\label{eqn:pctile-def}
\end{align}
where $p^{\ast} \equiv p \arg{\mu^{\ast} , \, E^{\ast}}$. This represents the probability that the true stellar parameters would be found at a point in ($\mu$, $E$)-space of lower posterior density. If ($\mu^{\ast}$, $E^{\ast}$) lies at the point of maximum posterior density, then $P \arg{p < p^{\ast}} = 1$. Conversely, if ($\mu^{\ast}$, $E^{\ast}$) lies in a region of vanishing probability density, then $P \arg{p < p^{\ast}} \approx 0$. This percentile is therefore similar to a cumulative distribution function, and is uniformly distributed between $0$ and $1$. A straightforward test of whether $P \arg{p < p^{\ast}} \sim U \arg{0 , \, 1}$ is to generate mock photometry for a large number of stars, to calculate the percentile for each simulated star, and to then bin the results. Each bin is expected to contain the same number of stars, with the precise number of stars in the bins determined by a multinomial distribution. We can therefore derive approximate confidence intervals for the number of stars that should fall into any given bin. Fig. \ref{fig:percentiles} shows this test for a set of 1000 simulated stars along a line of sight with Galactic coordinates $\ell = 90^{\circ}$, $b = 10^{\circ}$. The results are consistent with our expectations, indicating that our method recovers correct posterior densities for the simulated photometry.

\subsection{Mock Line of Sight}
\label{sec:mock-los}

\begin{figure*}[!htpb]
	\epsscale{1.0}
	\plotone{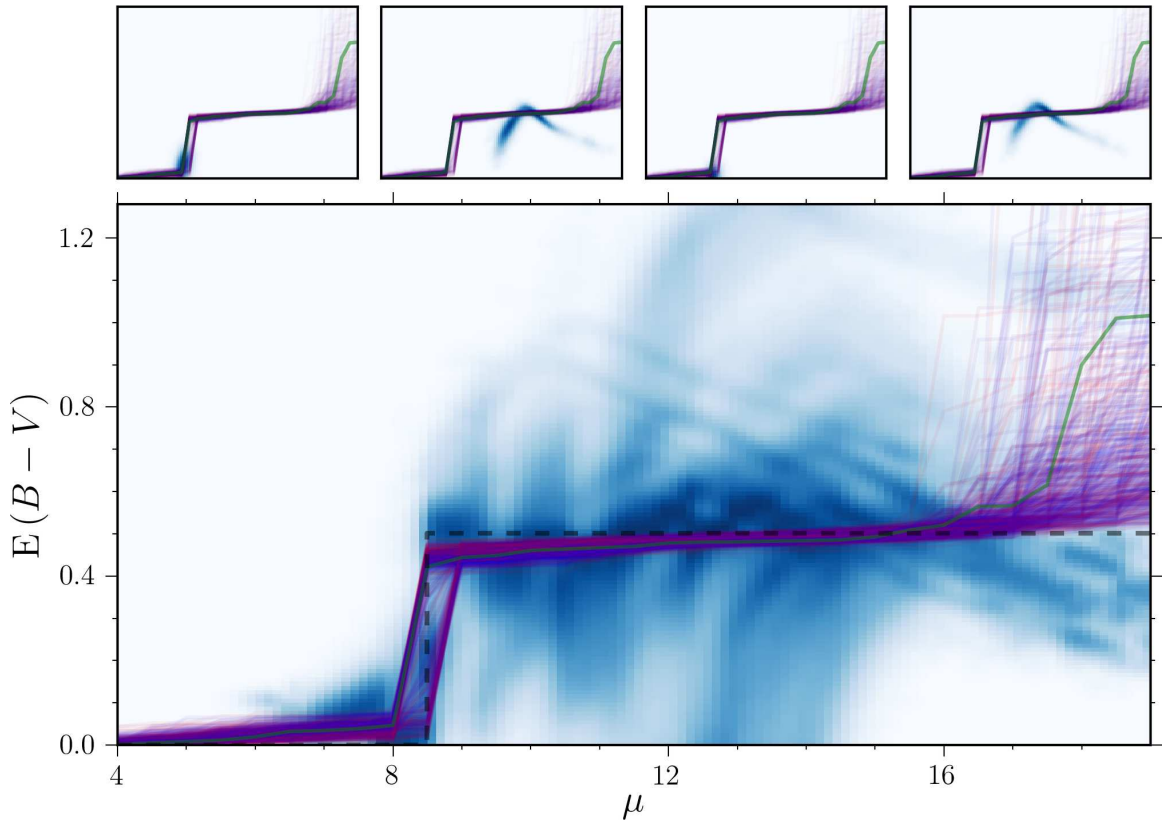}
	\caption{Recovery of the line-of-sight reddening profile from simulated photometry for 150 stars. The large panel shows the inferred posterior densities of the stars, stacked on top of one another. The contrast is stretched at each distance for purposes of visualization. This gives a picture of the information which is fed into the second stage of our analysis, in which we recover the reddening as a function of distance from the individual stellar probability densities (See Eq. \eqref{eqn:line-integral-product}). The stacked image, however, plays no direct role in this inference. The curves show possible reddening profiles, conditioned on the mock photometry. The green curve traces the most probable reddening profile. The remaining curves are colored according to the logarithm of their probability density, with blue denoting high probability and red denoting low probability. We recover a reddening profile similar to the original, which had a single cloud of depth $\EBV = 0.5$ at distance modulus $\mu = 8.5$ (shown as a dashed black line on the plot). The slight, gradual increase in inferred reddening beyond the cloud is due to the constraint that differential reddening in each bin be non-negative, and to the log-normal prior on differential reddening. \textit{A priori}, having no reddening away from the cloud is unlikely, and this is reflected in our inference. The upper four panels show individual stellar posterior density functions over the same domain, with the same reddening profiles overplotted. Each star is consistent with the range of possible reddening profiles. \label{fig:mock-los}}
\end{figure*}

Finally, we demonstrate that we are able to recover line-of-sight reddening profiles for simulated photometry. We first invent an arbitrary relationship $\EBV \arg{\mu}$ between distance and reddening. We add in low-level scatter to the distance-reddening relationship, as the reddening relation across one HEALPix pixel may vary. We then generate mock photometry for 150 stars along the line of sight. Following the procedure outlined in \S \ref{sec:los-formalism}, we use the simulated photometry to determine a posterior density in distance and reddening space for each star, and then combine the information from all of the stars to find the range of allowable reddening profiles. Our final product is thus a set of reddening profiles, drawn from the probability density over reddening profiles (Eq. \eqref{eqn:line-integral-product}). We parameterize the reddening as a piecewise-linear function in distance, and apply a weak log-normal prior to the differential reddening in each distance segment, as described in \S \ref{sec:los-fitting}. The results for one simulated line of sight, shown in Fig. \ref{fig:mock-los}, demonstrate that we are able to correctly infer the reddening profile for mock data. The method produces the best results at distances where there are many stars. Nearby and at large distances, where there are comparatively few stars to constrain the reddening profile, the uncertainties in reddening can become very large. However, at intermediate distances ($\mu \sim 10$ to $\sim 15$ for typical lines of sight, corresponding to $1$ to $10 \kpc$), the fit produces uncertainties on the order of $\Delta \EBV \sim 0.05 \, \mathrm{mag}$, consistent with the intrinsic scatter in reddening which we introduce into the simulated photometry.

\section{Comparison with Data}
\label{sec:comparison-with-data}

\subsection{Colors}
\label{sec:color-comparison}

We compare our model colors to PS1 stellar photometry from low-extinction regions at high Galactic latitudes. It is important to choose low-extinction regions, so that assumptions about the reddening law and what percentage of the total dust column is in front of each star play only a minor role. This allows us to obtain a comparison between the intrinsic colors in our model and of real stars. We de-redden the stellar colors assuming that they are behind the full dust column predicted by \citet[SFD]{Schlegel1998}.

\begin{figure*}[!htpb]
	\epsscale{0.7}
	\plotone{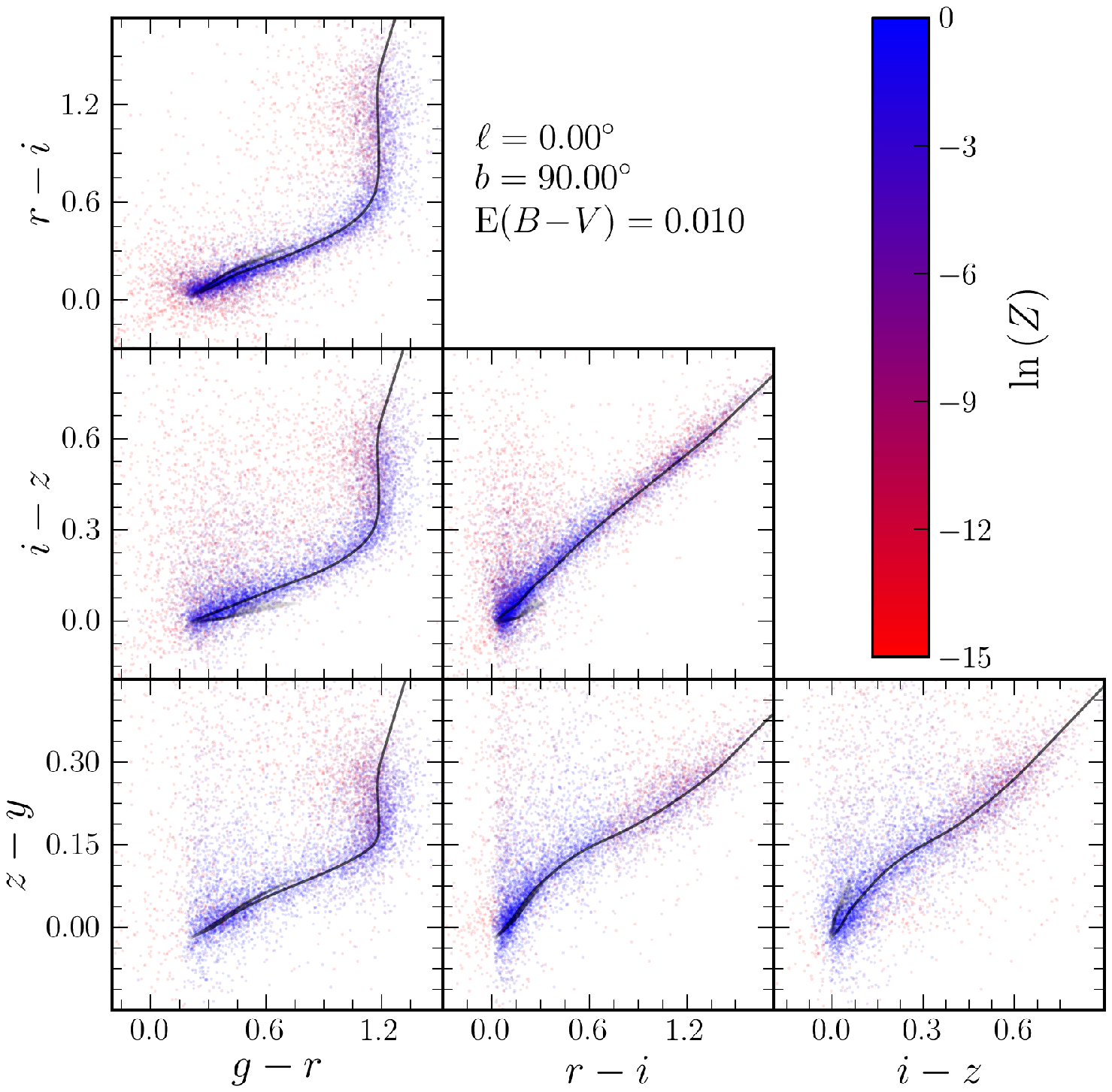}
	\caption{Comparison of PS1 stellar colors in the vicinity of the North Galactic Pole with our model colors. Each object is colored according to the evidence $Z$ we compute. Objects represented by red dots have a low probability of being drawn from our stellar model, and are rejected for the line-of-sight reddening determination. The solid black line traces our model stellar colors. Our main-sequence model colors do not depend on metallicity, while the model colors for the giant branch have a slight metallicity dependence. \label{fig:color-color-comparison}}
\end{figure*}

The results for the North Galactic Pole are shown in Fig. \ref{fig:color-color-comparison}. We compute an evidence for each star, as described in \S \ref{sec:evidence-outliers}. As expected, objects which lie far from the stellar locus in color-color space tend to have lower evidence. This helps us to reject objects that are either not stars, are not included in our stellar model (e.g., young and BHB stars), or that have particularly bad photometry. In the window shown here, 15\% of the detected objects would fail an evidence cut of $\ln Z > \ln Z_{\mathrm{max}} - 20$. These objects tend to have problematic photometry for which the PS1 pipeline may have produced inaccurate results, though some are variables, quasars, and unrecognized galaxies, which our technique is not designed to handle. Our line-of-sight reddening inferences are not strongly dependent on the choice of the evidence threshold.

\subsection{Distances}
\label{sec:distance-comparison}

\begin{figure*}[!htpb]
	\epsscale{1.0}
	\plotone{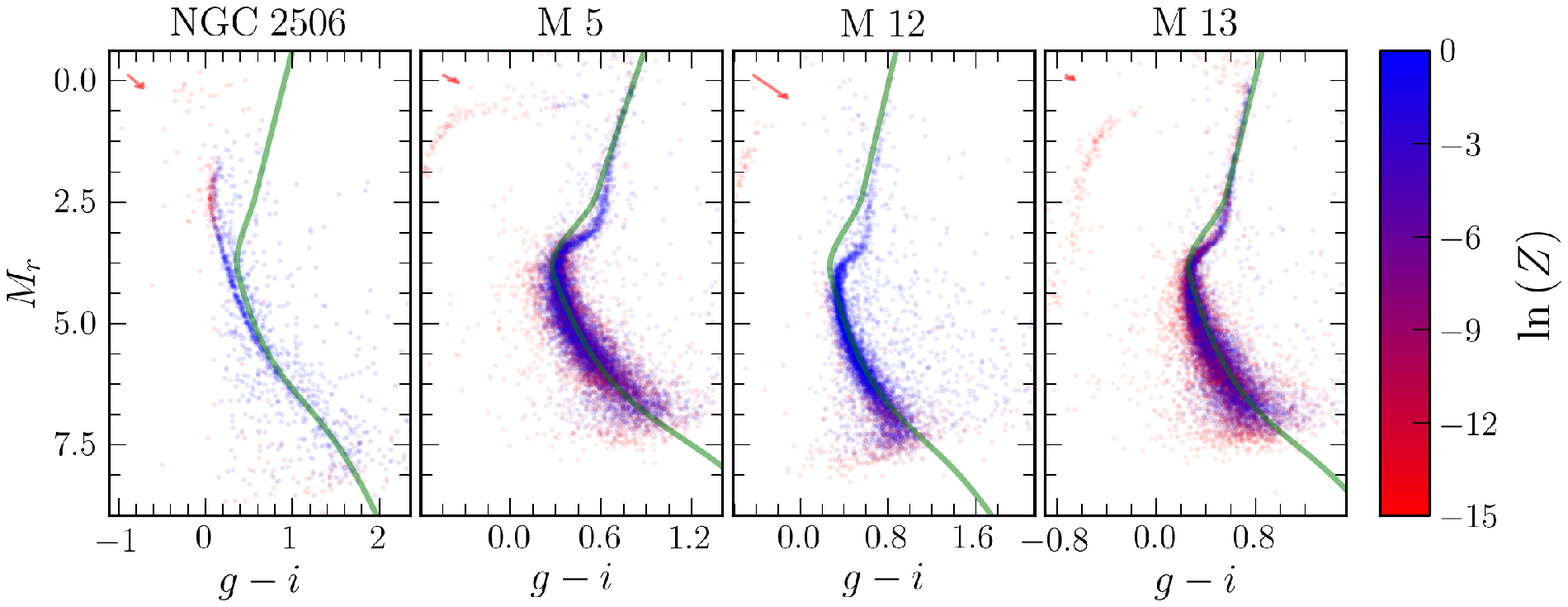}
	\caption{PS1 color-magnitude diagrams of three globular and one open cluster. For each cluster, the model isochrone with the catalog metallicity of the cluster is overplotted. The stellar photometry has been de-reddened and shifted by the catalog distance modulus to produce absolute magnitudes. The reddening vector is plotted in the top left corner of each panel in red for reference. Each star is colored by its evidence, with red stars unlikely to be drawn from our stellar model. In particular, stars which are blueward of the the main-sequence turnoff, which are bluer than any star in our template library, have low evidence. \label{fig:clusters}}
\end{figure*}

Correct distance determination requires not only correct model colors, but correct absolute magnitudes. We therefore compare our stellar models to globular and open clusters. In Fig. \ref{fig:clusters}, we compare our model magnitudes to photometry from four globular and open clusters.

We find that the model absolute magnitudes match the main sequence. The model magnitudes are unreliable past the main sequence turnoff, particularly for younger clusters. Our model magnitudes trace the giant branch of intermediate-age clusters---typical of the age of most stars in the Galaxy---somewhat better. Nevertheless, we expect that most of the information in our dust maps will come from the main sequence, where distance and reddening estimates are better constrained. Massive young blue stars and blue horizontal branch stars, which are not included in the stellar templates, are found to have low evidence, allowing them to be identified and excluded from the line-of-sight dust inference. Inclusion of age-dependent stellar models, and therefore more reliable distance and reddening determinations for the most massive stars, could potentially increase the distance to which our dust maps are reliable, and is an important direction for future work.

\subsection{Reddenings}
\label{sec:reddening-comparison}

In order to test the accuracy of our reddening inferences for individual stars, we compare our photometric reddenings to independently measured reddenings for a sample of stars. The Sloan Extension for Galactic Understanding and Exploration (SEGUE) \citep{Yanny2009}, part of SDSS-II, provides a convenient set of stars for which one can independently determine reddening. The SEGUE survey obtained moderate-resolution spectroscopy for 240,000 stars with SDSS photometry. Whereas most of the SDSS footprint is at low reddening, some of the SEGUE targets are at moderate reddening, up to $\sim 1 \, \mathrm{mag}$ in $\EBV$. The SEGUE Stellar Parameter Pipeline (SSPP) fits an atmospheric model to each star to derive the temperature, metallicity, and gravity of the star, as well as other parameters \citep{Lee2008, Lee2008a, AllendePrieto2008}.  These stellar parameters were used by \citet{Schlafly2011} to predict the intrinsic colors of stars, and to study the effect of reddening by attributing the differences between the observed and intrinsic colors to dust.  We use the reddening estimates of \citet{Schlafly2011} in the four independent SDSS colors, in concert with their recommended $R_V=3.1$ reddening vector \citep{Fitzpatrick1999}, to estimate $\EBV$ to each star. The details of deriving the reddening based on SEGUE-determined intrinsic colors and SDSS photometric colors are described in Appendix \ref{sec:app-SEGUE-reddenings}.

\begin{figure*}[!htpb]
	\epsscale{1.0}
	\plottwo{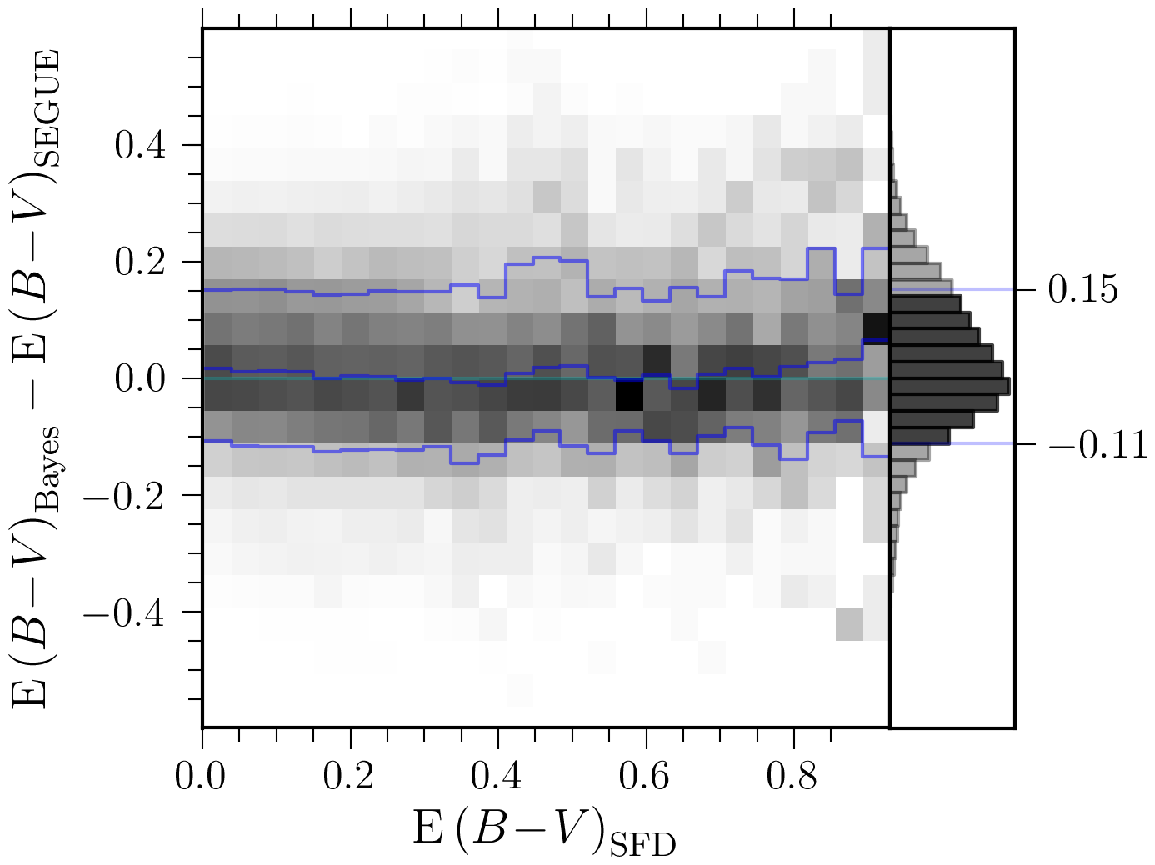}{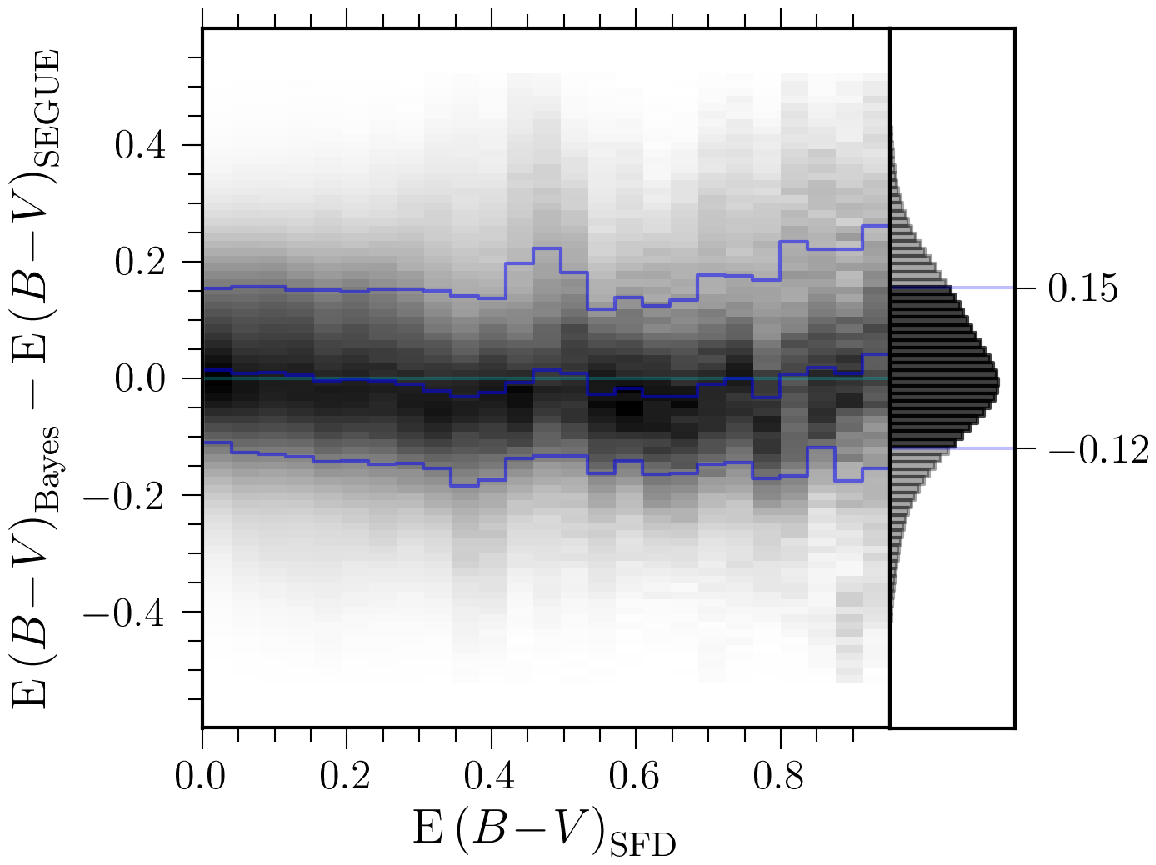}
	\caption{Histogram of the difference between the mode of the Bayesian reddening inference and the SEGUE-derived reddening, as a function of SFD reddening (See \S \ref{sec:reddening-comparison}). The blue envelopes mark the 15.87 and 84.13 percentiles of the residuals, equivalent to one standard deviation for a normal distribution, while the central blue curve marks the median of the residuals. The left panel compares the mode of the Bayesian posterior density functions with the mean of the SEGUE-derived reddenings. The right panel compares random samples drawn from the Bayesian posteriors with random samples drawn from the SEGUE-derived posteriors, which are Gaussian. \label{fig:SEGUE-comp}}
\end{figure*}

We use the same sample of SEGUE targets as \citet{Schlafly2011}. This sample excludes objects targeted as white dwarfs, and also removes M dwarfs, for which the stellar parameters are less reliable. We also require that each SEGUE target have a PS1 counterpart. Distances and reddenings to each star are then inferred as described in \S\ref{sec:likelihoods}, and compared to the SEGUE-determined reddenings.  For this comparison, we allow our photometric reddening estimates to be negative, for consistency with the SEGUE-derived reddenings. We save 100 reddening samples from the Markov chain for each star, as well as the maximum-posterior density reddening. In the left panel of Fig. \ref{fig:SEGUE-comp}, we compare the maximum-posterior density Bayesian reddening estimate with the SEGUE-derived reddening for 200,000 stars. We bin the stars by the reddening expected from the dust maps of \citet[SFD]{Schlegel1998}, and plot a histogram of the difference in the two reddening measures in each bin. We place the SFD reddening on the $x$-axis because it is a good proxy for reddening and is independent of both of the two reddening estimates we wish to compare, while placing either the SEGUE-derived reddening or the Bayesian reddening along the $x$-axis can introduce spurious trends in the resulting comparison. We find that the mode of our reddening estimates is unbiased over a range of $\EBV = 0$ to $1$, above which there are too few stars in the SEGUE sample to extend the comparison. The scatter in the difference between the two reddening estimates is approximately $0.12 \, \mathrm{mag}$ in \EBV, with the overall estimate being unbiased to within $0.03 \, \mathrm{mag}$.

Our Bayesian reddening and distance estimates assume that stars are drawn at random from the observable stars on each line of sight. The SEGUE survey, however, does not target stars at random, but instead targets only subsets of stars of particular interest. Moreover, the subset of SEGUE-observed stars for which we have reliable reddening estimates does not extend to the M-dwarfs, meaning that our sample of SEGUE-derived reddening estimates use only intrinsically blue stars. We consider intrinsically redder stars --- and therefore less reddened stars --- in our analysis than are actually present in our sample of SEGUE-observed stars. In order to simulate the effect of excluding M-dwarfs, we modify the luminosity function prior to assign zero probability for $M_{r} > 6$. When we account for this effect, the distribution of the difference between our and the SEGUE reddening estimates is unbiased. To illustrate this, instead of presenting single reddening difference for each star, we show 100 samples of the distribution of reddening differences between our Bayesian reddening estimate and the SEGUE-derived reddening estimate. The resulting residuals are shown as a function of SFD reddening in the right panel of Fig. \ref{fig:SEGUE-comp}. The residuals are unbiased at the $0.01 \, \mathrm{mag}$ level, with a scatter of $\EBV = 0.13 \, \mathrm{mag}$. This result is indicative of the accuracy we achieve for high-signal-to-noise detections, as most SEGUE targets are well above the detection limit in PS1.

\section{Conclusion}
\label{sec:conclusion}

We have presented a general method for deriving a three-dimensional map of Galactic reddening from stellar photometry. Our technique is based on grouping stars into pixels, determining the joint posterior of distance and reddening for each star, and then determining the most probable reddening-distance relation in each pixel. We have shown that this method correctly recovers the distance--reddening relationship for simulated lines of sight. We have additionally shown by comparison with SEGUE-derived reddenings that for high-SNR detections, our Bayesian reddening estimates are unbiased at the $0.01 \, \mathrm{mag}$ level, with a scatter of $\sim 0.13 \, \mathrm{mag}$ in $\EBV$. Based on comparisons with mock catalogs, in highly-reddened regions of the Galaxy, our distance inferences have typical uncertainties of $+$47\% on the high end, and $-$21\% on the low end. In high-Galactic latitude regions with low reddening, our distances have typical uncertainties of $+$52\% on the high end, and $-$38\% on the low end. These uncertainties may be reduced by feeding back information on reddening as a function of distance, derived from all the stars along the line of sight. A subsequent paper will present the results of applying the techniques developed here to construct a 3D reddening map covering the $\delta > -30^{\circ}$ sky.

In addition to determining the dust density in the nearby Galaxy, our method can be used to determine the distribution of stars in the Galactic plane. Earlier optical studies of the distribution of stars in the Galaxy traditionally consider only high-latitude stars, where the correction for dust extinction is straightforward \citep[e.g.,][]{Juric2008}. Infrared surveys of the plane are less sensitive to dust extinction, but their wavelength coverage also makes them less sensitive to intrinsic stellar type, rendering photometric distances uncertain. Our technique provides distances to stars throughout the Galactic plane, enabling future studies of the distribution of stars in the disk.

The technique described in this paper is not limited to PS1 photometry. Inclusion of information from the 2MASS $J$, $H$ and $K_{s}$ bands, WISE bands, as well as SDSS $u$-band photometry will improve our distance and reddening estimates. In addition, kinematic information, such as proper motion, may be incorporated into our framework in order to allow a more precise determination of stellar distances.

Upcoming surveys will also dramatically enhance our ability to measure the distances and reddenings to stars in the Galaxy. The LSST \citep{Ivezic2008a} will provide deeper photometry spanning a similar set of filters as those used in SDSS and Pan-STARRS 1, providing photometry for the sky south of $\delta < +34.5^{\circ}$. In the nearer future, the Dark Energy Survey (DES) will survey a complementary $5000 \, \mathrm{deg}^{2}$ of sky to Pan-STARRS 1, in a similar filter set \citep{DES2005}. The Gaia mission \citep{Lindegren1994}, meanwhile, will provide multiband photometry and low-resolution spectroscopy alongside parallax distance measurements and proper motions for one billion stars. Gaia's parallax distances, in particular, will break many of the degeneracies in our model for $r_{\mathrm{P1}} \lesssim 20$ stars, while its proper motions will aid in inferring the population each star belongs to. These new datasets will increase the power of our method to determine Galactic reddening and structure.

The Pan-STARRS1 Surveys (PS1) have been made possible through contributions of the Institute for Astronomy, the University of Hawaii, the Pan-STARRS Project Office, the Max-Planck Society and its participating institutes, the Max Planck Institute for Astronomy, Heidelberg and the Max Planck Institute for Extraterrestrial Physics, Garching, The Johns Hopkins University, Durham University, the University of Edinburgh, Queen’s University Belfast, the Harvard-Smithsonian Center for Astrophysics, the Las Cumbres Observatory Global Telescope Network Incorporated, the National Central University of Taiwan, the Space Telescope Science Institute, the National Aeronautics and Space Administration under Grant No. NNX08AR22G issued through the Planetary Science Division of the NASA Science Mission Directorate, the National Science Foundation under Grant No. AST-1238877, the University of Maryland, and Eotvos Lorand University (ELTE). Gregory M. Green and Douglas P. Finkbeiner are partially supported by NSF grant AST-1312891. The computations in this paper were run on the Odyssey cluster supported by the FAS Science Division Research Computing Group at Harvard University.

\appendix

\section{Harmonic Mean Estimate of the Bayesian Evidence}
\label{sec:app-harmonic-mean}

The harmonic mean approximation, developed in \citet{Gelfand1994}, allows one to compute the Bayesian evidence using samples returned from a Markov Chain Monte Carlo simulation. For a model with parameters $\theta$ and data $D$, Bayes' rule tells us that
\begin{align}
	p \arg{\theta | D} = \frac{p \arg{D | \theta} p \arg{\theta}}{p \arg{D}} \, .
\end{align}
We wish to compute the evidence $p \arg{D}$, often denoted by $Z$. Multiplying each side of the above by an arbitrary function $\phi \arg{\theta}$, rearranging terms and taking the integral over all $\theta$,
\begin{align}
	\frac{1}{p \arg{D}} \int \! \d \theta \, \phi \arg{\theta} &= \int \! \d \theta \, p \arg{\theta | D} \, \frac{\phi \arg{\theta}}{p \arg{D | \theta} p \arg{\theta}} \, .
\end{align}
The r.h.s. is simply the expectation value of
\begin{align}
	\frac{\phi \arg{\theta}}{p \arg{D | \theta} p \arg{\theta}}
\end{align}
for samples drawn from the posterior density $p \arg{\theta | D}$. This is convenient, since MCMC methods draw a set of samples from the distribution $p \arg{\theta | D}$. If the integral of $\phi \arg{\theta}$ is normalized to unity, then
\begin{align}
	\frac{1}{p \arg{D}} &\equiv \frac{1}{Z} \approx \left< \frac{\phi \arg{\theta}}{p \arg{D | \theta} p \arg{\theta}} \right>_{\! \mathrm{chain}} .
\end{align}
This estimate has finite variance as long as $\phi$ has steeper wings than $p \arg{\theta | D}$ \citep{Robert2009}. We choose $\phi$ to be constant within an ellipse centered on a point of high density within the chain, and zero outside \citep{Robert2009}. The ellipse is aligned with the principle axes of the covariance matrix, in order to ensure that it contains only well sampled regions of parameter space.

We find a point of high density by first centering an ellipse on a random point in the chain. Because the points in the chain are sampled proportionately to the posterior probability density, this point is already likely to lie in a well-sampled region of parameter space. We then find the mean of the points in the chain falling in the ellipse, and move the center of the ellipse to that position in parameter space. One can iterate this procedure several times to settle into a densely sampled region of parameter space. The size of the ellipse we use to define $\phi \arg{\theta}$ is chosen such that a preset fraction of samples in the chain are enclosed. For our calculations, we iterate five times to find a dense region of parameter space, and scale the ellipse such that it contains 5\% of samples.

\section{SEGUE-Derived Reddenings}
\label{sec:app-SEGUE-reddenings}
Here, we review a method for calculating stellar reddenings on the basis of SDSS photometry and SEGUE predicted intrinsic stellar colors. As explained in \citet{Schlafly2011}, since the extinction in an individual band $X$ is given by
\begin{align}
	A_{X} &= R_{X} \EBV \, ,
\end{align}
colors transform as
\begin{align}
	\mathrm{E} \! \left( X \! - \! Y \right)
	&= A_{Y} - A_{X}
	= \left( R_{Y} - R_{X} \right) \EBV \, .
	\label{eqn:single-color-reddening-estimate}
\end{align}
If we have only one color, $X - Y$, we can therefore estimate reddening as
\begin{align}
	\EBV &=
	\frac{ \mathrm{E} \! \left( X \! - \! Y \right) }{ R_{Y} - R_{X} } \, .
\end{align}

Our goal is to extend this formula to allow the use of multiple colors, possibly with strong covariance. Since \citet{Schlafly2011} only predict the colors of stars, and not their overall magnitudes, we work in color space. Denote the intrinsic stellar colors as $\vec{c}_{i}$, and the reddened colors as $\vec{c}_{r}$.

In a multidimensional color space, Eq. \eqref{eqn:single-color-reddening-estimate} becomes
\begin{align}
	\vec{c}_{r} - \vec{c}_{i} = \vec{R} \, \EBV \, ,
	\label{eqn:SEGUE-true-color-relation}
\end{align}
where $\vec{R}$ has one component per color $X - Y$, given by $R_{XY} \equiv R_{Y} - R_{X}$. The estimated intrinsic colors and observed reddened colors are Gaussian random variables, with covariances $\Sigma_{i}$ and $\Sigma_{r}$, respectively. Denote the estimated intrinsic colors as $\vec{c}_{i}^{\, \prime}$, and the observed reddened colors as $\vec{c}_{r}^{\, \prime}$. The likelihood of these two quantities taking on a particular set of values is given by
\begin{align}
	p \arg{ \vec{c}_{i}^{\, \prime} , \, \vec{c}_{r}^{\, \prime} \, | \, \EBV , \, \vec{c}_{r} }
	&= \mathcal{N} \arg{ \vec{c}_{r}^{\, \prime} \, | \, \vec{c}_{r} , \, \Sigma_{r} } \mathcal{N} \arg{ \vec{c}_{i}^{\, \prime} \, | \, \vec{c}_{i} , \, \Sigma_{i} } \\
	&= \mathcal{N} \arg{ \vec{c}_{r}^{\, \prime} \, | \, \vec{c}_{r} , \, \Sigma_{r} } \mathcal{N} \arg{ \vec{c}_{i}^{\, \prime} \, | \, \vec{c}_{r} - \vec{R} \, \EBV , \, \Sigma_{i} } \, .
\end{align}
In the second step, we have replaced $\vec{c}_{i}$ using Eq. \eqref{eqn:SEGUE-true-color-relation}. Using the symmetry of the Gaussian distribution,
\begin{align}
	p \arg{ \vec{c}_{i}^{\, \prime} , \, \vec{c}_{r}^{\, \prime} \, | \, \EBV , \, \vec{c}_{r} }
	&= \mathcal{N} \arg{ \vec{c}_{r} \, | \, \vec{c}_{r}^{\, \prime} , \, \Sigma_{r} } \mathcal{N} \arg{ \vec{c}_{r} - \vec{R} \, \EBV \, | \, \vec{c}_{i}^{\, \prime} , \, \Sigma_{i} } \\
	&= \mathcal{N} \arg{ \vec{c}_{r} \, | \, \vec{c}_{r}^{\, \prime} , \, \Sigma_{r} } \mathcal{N} \arg{ \vec{R} \, \EBV - \vec{c}_{r} \, | \, -\vec{c}_{i}^{\, \prime} , \, \Sigma_{i} } \, .
\end{align}
If we assume a flat prior on $\EBV$ and $\vec{c}_{r}$, then the above is proportional to the posterior probability density $p \arg{ \EBV , \, \vec{c}_{r} \, | \, \vec{c}_{i}^{\, \prime} , \, \vec{c}_{r}^{\, \prime} }$. We could, in practice, frame the question in terms of the intrinsic colors $\vec{c}_{i}$, and put priors on them based on our Galactic and stellar model, but we wish to avoid tying our SEGUE-derived reddenings in any way to our Bayesian photometric reddening estimates. Now, integrating over $\vec{c}_{r}$, we obtain a convolution of two Gaussians, which is itself a Gaussian distribution:
\begin{align}
	p \arg{ \EBV \, | \, \vec{c}_{i}^{\, \prime} , \, \vec{c}_{r}^{\, \prime} }
	&\propto \int \d \vec{c}_{r} \, \mathcal{N} \arg{ \vec{c}_{r} \, | \, \vec{c}_{r}^{\, \prime} , \, \Sigma_{r} } \mathcal{N} \arg{ \vec{R} \, \EBV - \vec{c}_{r} \, | \, -\vec{c}_{i}^{\, \prime} , \, \Sigma_{i} } \\
	&\hspace{0.2cm} = \mathcal{N} \arg{ \vec{R} \, \EBV \, | \, \vec{c}_{r}^{\, \prime} - \vec{c}_{i}^{\, \prime} , \, \Sigma_{r} + \Sigma_{i} }
\end{align}
The probability density function of $\EBV$ is thus a ray taken through a multivariate Gaussian. It can be shown that the above is also Gaussian, with mean and standard deviation given by
\begin{align}
	\left< \EBV \right> &= \frac{
	\left( \vec{c}_{r}^{\, \prime} - \vec{c}_{i}^{\, \prime} \right)^{T} \left( \Sigma_{r} + \Sigma_{i} \right)^{-1} \vec{R}
	}{
	\vec{R}^{T} \left( \Sigma_{r} + \Sigma_{i} \right)^{-1} \vec{R}
	}
	\, , \\
	\sigma_{\EBV}^{2} &= \left[ \vec{R}^{\, T} \left( \Sigma_{r} + \Sigma_{i} \right)^{-1} \vec{R} \, \right]^{-1} \, .
\end{align}
We plug intrinsic stellar colors appropriate for the SSPP stellar parameters into $\vec{c}_{r}^{\, \prime}$, and use the observed SDSS colors for $\vec{c}_{i}^{\prime}$.

One final note is that \citet{Schlafly2011} used the SSPP stellar types to derive estimates of the mean and covariance of the magnitudes, rather than the colors. We will now show how to obtain the covariance of the colors from the covariance matrix of the magnitudes. Denote the covariance matrix of the magnitudes as $\Sigma_{ij}$, where $i$ and $j$ label passbands. Label the color $m_{i} - m_{j}$ as $c_{ij}$. We then call the covariance matrix of the colors $c_{ij}$ and $c_{k \ell}$ $\Sigma_{ij, k \ell}^{\, \prime}$. By expanding out
\begin{align}
	\Sigma_{ij, k \ell}^{\, \prime} &= \left< c_{ij} c_{k \ell} \right> - \left< c_{ij} \right> \left< c_{k \ell} \right>
\end{align}
in terms of magnitudes, one obtains
\begin{align}
	\Sigma_{ij, k \ell}^{\, \prime} &= \Sigma_{ik} - \Sigma_{i \ell} - \Sigma_{jk} + \Sigma_{j \ell} \, .
\end{align}
The common choice of colors is to set the $i^{\mathrm{th}}$ color to $m_{i} - m_{i+1}$. Plugging $j = i+1$ and $\ell = k + 1$ into the above, we find that the covariance of the $i^{\mathrm{th}}$ color with the $k^{\mathrm{th}}$ color is given by
\begin{align}
	\Sigma_{i, k}^{\, \prime} &= \Sigma_{i, k} - \Sigma_{i, k+1} - \Sigma_{i+1, k} + \Sigma_{i+1, k+1} \, .
\end{align}

\bibliographystyle{apj}

\end{document}